\documentclass[a4paper,fleqn]{cas-dc}

\usepackage[authoryear]{natbib}

\def\tsc#1{\csdef{#1}{\textsc{\lowercase{#1}}\xspace}}
\tsc{WGM}
\tsc{QE}
\tsc{EP}
\tsc{PMS}
\tsc{BEC}
\tsc{DE}

\def \degmark{^\circ}

\def \arcmin {\hbox{$^\prime$}}
\def \arcsec {\hbox{$^{\prime\prime}$}}

\def \gray {$\gamma$-ray}

\begin{document}
\let\WriteBookmarks\relax
\def\floatpagepagefraction{1}
\def\textpagefraction{.001}
\shorttitle{The ASTRI Mini-Array}
\shortauthors{S Scuderi et~al.}

\title [mode = title]{The ASTRI Mini-Array of Cherenkov Telescopes at the Observatorio del Teide}                      
%
%

\author[1]{S. Scuderi}[type=editor,
                        ]
\cormark[1]
\ead{salvatore.scuderi@inaf.it}
\author[1]{A. Giuliani}
\author[2]{G. Pareschi}
\author[3]{G. Tosti}
\author[6]{O. Catalano}
\author[16]{E. Amato}
\author[8]{L.A. Antonelli}
\author[13]{J. {Becerra Gonz\`ales}}
\author[4]{G. Bellassai}
\author[8, 21]{C. Bigongiari}
\author[6]{B. Biondo}
\author[14]{M. B\"ottcher}
\author[4]{G. Bonanno}
\author[2]{G. Bonnoli}
\author[4]{P. Bruno}
\author[5]{A. Bulgarelli}
\author[6]{R. Canestrari}
\author[6]{M. Capalbi}
\author[1]{P. Caraveo}
\author[11]{M. Cardillo}
\author[5]{V. Conforti}
\author[6]{G. Contino}
\author[6]{M. Corpora}
\author[4]{A. Costa}
\author[6]{G. Cusumano}
\author[6]{A. D'A\`i}
\author[12]{E. {de Gouveia Dal Pino}}
\author[2]{R. {Della Ceca}}
\author[15]{E. {Escribano Rodriguez}}
\author[19]{D.  {Falceta-Gon\c{c}alves}}
\author[12]{C. Fermino}
\author[10,7]{M. Fiori}
\author[5]{V. Fioretti}
\author[1]{M. Fiorini}
\author[8]{S. Gallozzi}
\author[6]{C. Gargano}
\author[4]{S. Garozzo}
\author[3]{S. Germani}
\author[15]{A. Ghedina}
\author[5]{F. Gianotti}
\author[6]{S. Giarrusso}
\author[6, 12]{R. Gimenes}
\author[4]{V. Giordano}
\author[4]{A. Grillo}
\author[15]{C. {Grivel Gelly}}
\author[6]{D. Impiombato}
\author[4]{F. Incardona}
\author[1]{S. Incorvaia}
\author[2]{S. Iovenitti}
\author[6]{A. {La Barbera}}
\author[1]{N. {La Palombara}}
\author[6]{V. {La Parola}}
\author[8]{A. Lamastra}
\author[7]{L. Lessio}
\author[4]{G. Leto}
\author[6]{F. {Lo Gerfo}}
\author[15]{M. Lodi}
\author[8]{S. Lombardi}
\author[18]{F. Longo}
\author[8, 21]{F. Lucarelli}
\author[6]{M.C. Maccarone}
\author[4]{D. Marano}
\author[4]{E. Martinetti}
\author[1]{S. Mereghetti}
\author[4]{A. Miccich\'e}
\author[2]{R. Millul}
\author[6]{T. Mineo}
\author[6]{D. Mollica}
\author[17]{G. Morlino}
\author[9]{A. Morselli}
\author[10,7]{G. Naletto}
\author[20]{G. Nicotra}
\author[6]{A. Pagliaro}
\author[5]{N. Parmiggiani}
\author[11]{G. Piano}
\author[6]{F. Pintore}
\author[15]{E. Poretti}
\author[17]{B. Olmi}
\author[5]{G. Rodeghiero}
\author[9]{G. {Rodriguez Fernandez}}
\author[2]{P. Romano}
\author[4]{G. Romeo}
\author[5]{F. Russo}
\author[6]{P. Sangiorgi}
\author[8]{F.G. Saturni}
\author[2]{J.H. Schwarz}
\author[4]{E. Sciacca}
\author[2]{G. Sironi}
\author[6]{G. Sottile}
\author[8]{A. Stamerra}
\author[2]{G. Tagliaferri}
\author[8]{V. Testa}
\author[4]{G. Umana}
\author[1]{M. Uslenghi}
\author[2]{S. Vercellone}
\author[7]{L. Zampieri}
\author[4]{R. {Zanmar Sanchez}}

\address[1]{INAF, IASF Milano, Via Alfonso Corti 12, I-20133 Milano, Italy}
\address[2]{INAF, OA Brera, Via Brera, 28, I-20121 Milano, Italy}
\address[3]{Università di Perugia, Via Alessandro Pascoli, I-06123 Perugia, Italy}
\address[4]{INAF, OA Catania, Via Santa Sofia, 78, I-95123 Catania, Italy}
\address[5]{INAF, OAS Bologna, Via Piero Gobetti, 93/3, I-40129 Bologna, Italy}
\address[6]{INAF, IASF Palermo, Via Ugo la Malfa, 153, I-90146 Palermo, Italy}
\address[7]{INAF, OA Padova, Vicolo Osservatorio 5, I-35122, Padova, Italy}
\address[8]{INAF, OA Roma, Via Frascati 33, I-00078 Monte Porzio Catone, Italy}
\address[9]{INFN, Sezione di Roma 2 c/o Dipartimento di Fisica, II Università di Roma ``Tor Vergata'', Via della Ricerca Scientifica 1, I-00133, Roma, Italy}
\address[10]{Dipartimento di Fisica ed Astronomia, Università di Padova, Via F. Marzolo 8, I-35131 Padova, Italy}
\address[11]{INAF, IAPS Roma, Via Fosso del Cavaliere 100, I-00133 Roma, Italy}
\address[12]{Instituto de Astronomia, Geofísica e Ciências Atmosféricas, Universidade de São Paulo, R. do Matão 1226, Cd. Universitária, 05508-090 São Paulo, Brazil}
\address[13]{Instituto de Astrofísica de Canarias \& Universidad de La Laguna, C. Vía Láctea, s/n, 38205 San Cristóbal de La Laguna, Santa Cruz de Tenerife, Spain}
\address[14]{Centre for Space Research, North-West University, Potchefstroom, South Africa}
\address[15]{Fundacion Galileo Galilei-INAF, Rambla José Ana Fernández Pérez, 7, 38712 Breña Baja, TF - Spain}
\address[16]{INAF, OA Arcetri, Largo Enrico Fermi, 5, 50125 Firenze, Italy}
\address[17]{INAF, OA Palermo, Piazza del Parlamento, 1, 90134 Palermo, Italy}
\address[18]{Università di Trieste \& INFN Trieste, Galleria Padriciano, 99, 34149 Trieste TS, Italy}
\address[19]{Escola de Artes, Ciências e Humanidades, Universidade de São Paulo, Av Arlindo Bettio, 1000, São Paulo, 03828-000 - Brasil}
\address[20]{INAF, IRA Via Piero Gobetti, 101, 40129 Bologna, Italy}
\address[21]{ASI - Space Science Data Center, Via del Politecnico s.n.c., I-00133, Rome, Italy}


\cortext[cor1]{Corresponding author}


\begin{abstract}
The ASTRI Mini-Array (MA) is an INAF project to build and operate a facility to study astronomical sources emitting at very high-energy in the TeV spectral band. The ASTRI MA consists of a group of nine innovative Imaging Atmospheric Cherenkov telescopes. The telescopes will be installed at the Teide Astronomical Observatory of the Instituto de Astrofisica de Canarias (IAC) in Tenerife (Canary Islands, Spain) on the basis of a host agreement with INAF. Thanks to its expected overall performance, better than those of current Cherenkov telescopes' arrays for energies above $\sim$5 TeV and up to 100 TeV and beyond, the ASTRI MA will represent an important instrument to perform deep observations of the Galactic and extra-Galactic sky at these energies. 
\end{abstract}



\begin{keywords}
Imaging Atmospheric Cherenkov Telescope \sep Very High-energy Gamma Rays \sep ASTRI \sep Schwarzschild-Couder telescopes
\end{keywords}

\maketitle
\tableofcontents

\section{Introduction}

\subsection{The ASTRI program}
The ASTRI program was born as a “flagship project” funded by the Italian Ministry of University and Scientific Research, led by the Italian National Institute for Astrophysics (INAF), and finalized to the technological development of the next generation of Imaging Atmospheric Cherenkov Telescopes (IACT) for ground-based gamma ray astronomy in the framework of the development of the CTA observatory. The acronym ASTRI stands for “Astronomia a Specchi a Tecnologia Replicante Italiana” which means Astronomy with mirrors built through Italian replica technology. The term was coined by Nanni Bignami (who was INAF president at that time and till the end was an enthusiastic supporter of the project) and is related to the production process of the sandwiched primary mirror segments based on a cold replication of thin glass sheets \citep{2013SPIE.8861E..03P}. 

The project involves more than 150 researchers of the INAF institutes in Bologna, Catania, Milan, Padua, Palermo, and Rome. The universities of Perugia, Padua, Catania, Genova and the Milan Polytechnic together with the INFN sections of Roma Tor Vergata and Perugia also participate to the project. The university of Sao Paulo in Brazil, the North Western University in South Africa and the Instituto de Astrofisica de Canarias (IAC) in Spain are the international partners of the project. Finally, several Italian and foreign industrial companies actively participate to the project. 

This paper will provide a detailed description of the ASTRI Mini-Array. In particular, we describe the site and the general architecture of the system. We also give a technical description of the main subsystems and their functions. Finally, we briefly describe the implementation and the science operation plan. This work (Paper I) has three companion papers. In \citep{2022JHEAp...S..XXXS}, Paper II, we describe the Core Science that ASTRI Mini-Array will address during the first years of operation. In \citep{2022JHEAp...A..XXXD}, Paper III, and \citep{2022JHEAp...F..XXXS}, Paper IV, we report the expected results on Galactic and extra-galactic sources, respectively, that the ASTRI Mini-Array will achieve during its observatory phase, i.e. after the completion of the core science observing period. 

\subsection{The ASTRI-Horn prototype}
The first phase of the project consisted in the design, realization and deployment of a technological demonstrator, an innovative end-to-end prototype of the 4 meters class telescopes that was named ASTRI-Horn in honour of Guido Horn D'Arturo, a precursor in the technique of segmented mirrors \citep{1936MmSAI...9..133H}. The main innovations of ASTRI-Horn telescope are the dual mirrors optical configuration based on a modified Schwarzschild - Couder design, \citep{2007APh....28...10V} and a compact Cherenkov camera based on Silicon Photo-Multipliers (SiPM) detectors and an innovative readout electronics \citep{2016SPIE.9906E..3DS},  \citep{2018SPIE10702E..37C}. The end-to-end philosophy implies that its functionality was to be proven using not only technical tests but also through the observations of gamma ray astrophysical sources. To implement the full chain of events that goes from the ``photon capture'' to the analysis of the data, it was necessary to develop not only the telescope and the Cherenkov camera but also internal and external calibration systems, hardware and control software, data reduction and data analysis software and the archive system.

The ASTRI-Horn telescope was inaugurated in 2014 at the M.G. Fracastoro station of the INAF – Catania Astrophysical Observatory, placed at an altitude of 1725 meters above the sea level inside the Etna regional park. The performance of the telescope and Cherenkov Camera have been extensively tested \citep{2016SPIE.9906E..19C}, \citep{2017SPIE10399E..04C}, \citep{2018SPIE10700E..5ZS}, \citep{2019SPIE11119E..1EG}, and \citep{2019SPIE11114E..0AC} proving the validity of design concept. The validation of the Schwarzschild-Couder optical concept was obtained in October 2016 \citep{2017A&A...608A..86G} while the detection of the first Cherenkov light was obtained in May 2017 and that of the Crab Nebula at energies larger than 3.5 TeV the year after \citep{2020A&A...634A..22L}. 

The task of the ASTRI-Horn telescope is not finished yet as it will be used as a test bench for the implementation of hardware and software for the continuation of the project. The telescope is currently undergoing maintenance, with the replacements of segments of the primary mirror, the refurbishment of the prototype camera and the maintenance of the active mirror control system, to be again operational at the beginning of Spring 2022. 

\subsection{The ASTRI Mini-Array: a new infrastructure in the IACT arrays context}
After the realization and validation of the ASTRI-Horn prototype telescope the project moved forward. The construction of three new telescopes, that became nine thanks to new funds made available from the Italian government, the University of Sao Paulo in Brasil and the North West University in South Africa, was foreseen since the beginning of the project. This set of nine telescopes has been named ASTRI Mini-Array. The telescopes were supposed to be pathfinders for the south site of the CTA observatory. However, the CTA timeline was not compatible with the ASTRI funds availability, so in 2019 INAF and the Instituto de Astrofisica des Canarias (IAC) came to an agreement to install the ASTRI Mini-Array at the Teide Astronomical Observatory, operated by IAC, in the Canary island of Tenerife. The ASTRI Mini-Array project will be also supported by the ``Fundación Galileo Galilei - INAF, Fundación Canaria'' (FGG). The FGG is a Spanish no-profit institution constituted by INAF whose aim is to manage and run the Telescopio Nazionale Galileo (\href{http://www.tng.iac.es}{www.tng.iac.es}) and to promote the astrophysical research in the Canary Islands on behalf of INAF. In the framework of the ASTRI Mini-Array the FGG will manage administrative activities and support all activities related to the implementation and operations of the ASTRI Mini-Array. 

The hosting agreement with IAC provides for at least 4+4 years of operations. During at least the first 3 years of operations the ASTRI Mini-Array will be run as an experiment and not as an observatory. After this initial period the ASTRI Mini-Array will gradually move towards an observatory model in which a fraction of the time will be assigned to scientific proposal going through a Time Allocation Committee procedure.

The ASTRI mini-array will be the largest facility of IACT arrays until the CTA observatory will start operations. Compared to currently operating IACT systems it will be more sensitive at energies larger than 1 TeV with 50 hours of observations (see Figure 9 in Vercellone et al, Paper II) and will extend the sensitivity up to 100 TeV and beyond, an almost never explored energy range by IACTs.
Furthermore, the ASTRI Mini-Array will operate when the present IACT systems, observing in a lower but partly overlapping energy range, will still be active, allowing direct comparison of scientific data (spectra, light-curves, integral fluxes). Also, fruitful synergies are clearly foreseen with HAWC and possibly with LHAASO, surveying a very large area of the northern sky, with pointed observations to characterize the morphology of extended sources detected at the highest VHE by these facilities. 

The ASTRI Mini-Array telescopes, based on the evolution of the ASTRI-Horn prototype, will implement a larger field of view (more than 10$\degmark$ in diameter) and will be equipped with the updated version of the compact ASTRICAM camera, based on Hamamatsu silicon photomultipliers and on the CITIROC-1A ASIC developed by Weeroc in collaboration with INAF for the read-out electronics.
The large field of view will allow to monitor simultaneously a few nearby sources during the same pointing. The combination of the sensitivity extended to 100 TeV and of the homogeneous performance across the FOV will allow us to study the emission from extended sources such as SNRs and PWNs at E > 10 TeV, and to investigate the presence of spectral cut-offs. The use of a SiPM-based camera will improve the duty cycle of the system allowing safe and effective operation with any level of moon condition as already demonstrated by FACT telescope \citep{2017ICRC...35..609D} and very recently by LHAASO \citep{2021EPJC...81..657A}. 

Even if gamma-ray astrophysics makes undoubtedly its core science, the ASTRI Mini-Array will also be capable of exploring other scientific topics. In particular:

\textit{Stellar Hambury-Brown intensity interferometry}: each of the telescopes of the ASTRI Mini-Array will be equipped with an intensity interferometry module. The Mini-Array layout with its very long baselines (hundreds of meters), will allow, in principle, to obtain angular resolutions as good as 50 micro-arcsec. With this level of resolution, it will be possible to reveal details on the surface and of the environment surrounding bright stars on the sky opening unprecedented frontiers in some of the major topics in stellar astrophysics.

\textit{Direct measurements of cosmic rays}: 99$\%$ of the observable component of the Cherenkov light is hadronic in nature. Even if the main challenge in detecting gamma-rays is to distinguish them from the much wider background of hadronic Cosmic Rays, this background, recorded during normal $\gamma$-ray observations, will be used to perform direct measurements and detailed studies on the Cosmic Rays. 

In summary, the ASTRI mini-array will allow to carry out seminal studies on both galactic and extragalactic sources, tackling frontier issues at the intersection of the fields of astrophysics, cosmology, particle physics and fundamental physics.

\begin{figure*}
	\centering
		\includegraphics[scale=1.4]{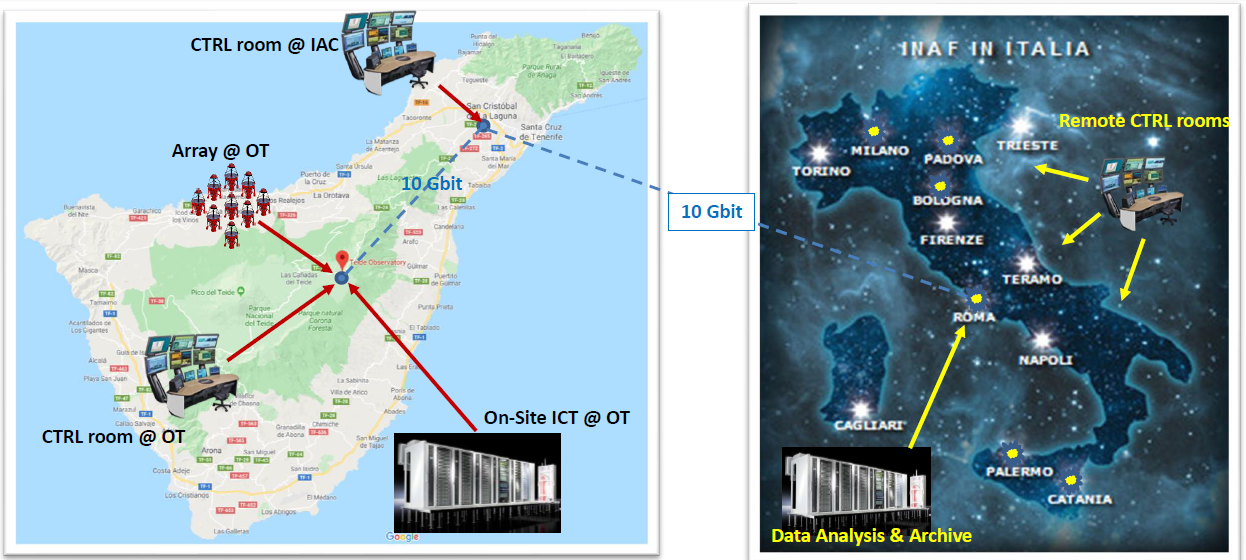}
	\caption{Locations of the ASTRI Mini-Array. Left panel: the positions of the ASTRI Mini-Array facilities in Tenerife. Right panel: the locations of INAF institutes with highlighted those involved in the ASTRI project. The dashed lines, labelled 10 Gbit, shows the network connection between the different locations}
	\label{FIG:Locations}
\end{figure*}

\section{Locations of the ASTRI Mini-Array}
The ASTRI Mini-Array will be distributed among different locations in Tenerife (Figure \ref{FIG:Locations}) and Italy:
\begin{itemize}
    \item The ASTRI Mini Array Site at the Teide Observatory will host the telescope array, auxiliary equipment (e.g. weather stations, LIDARs etc.) and site support facilities (e.g. control room, on-site data centre, power management system etc.). In particular, the on-site control room will be hosted at the THEMIS telescope and the on-site data centre at the Residencia of the Teide Observatory.
    \item The ASTRI Mini Array Support Site located at IAC premises in La Laguna, hosting the administrative services, the remote control room, and warehouse spaces. 
    \item Italy will host the off-site data centre at the INAF Astronomical Observatory in Rome and remote control rooms called Array Observations Centers (AOC) in the INAF institutes involved in the project. 
\end{itemize}

\section{The observational site of the ASTRI Mini-Array}
\subsection{The Teide Observatory}
The Canaries have two islands with world-class skies for astronomy that host two international observatories which together constitute the most important optical, infrared and gamma ray observatories in Europe: the “Observatorio del Teide” (OT) at Tenerife and the “Observatorio del Roque de los Muchachos” at La Palma (ORM). The observatory is operated by IAC.
The ASTRI Mini-Array site is located in the island of Tenerife (Canarias, Spain) at coordinates 28$\degmark{}$ 17$\arcmin{}$ 60.00$\arcsec{}$ N -16$\degmark{}$ 30$\arcmin{}$ 20.99$\arcsec{}$ W and at an altitude of approximately 2370 m. 
The temperature inversion layer in Canarias appears around 1300 m of altitude on average even if its altitude and thickness have a seasonal dependence, being higher and thinner during the winter (when it is located between 1350 m and 1850 m above the sea level, being only 350 m thick) and lower and thicker during the Summer (between 750 m and 1400 m above sea level, being about 550 m thick). The ASTRI Mini-Array site is permanently above this inversion layer with clean air and clear sky conditions that prevail all around the year.

\subsection{The layout of the ASTRI Mini-Array}
\begin{figure*}
	\centering
		\includegraphics[scale=0.9]{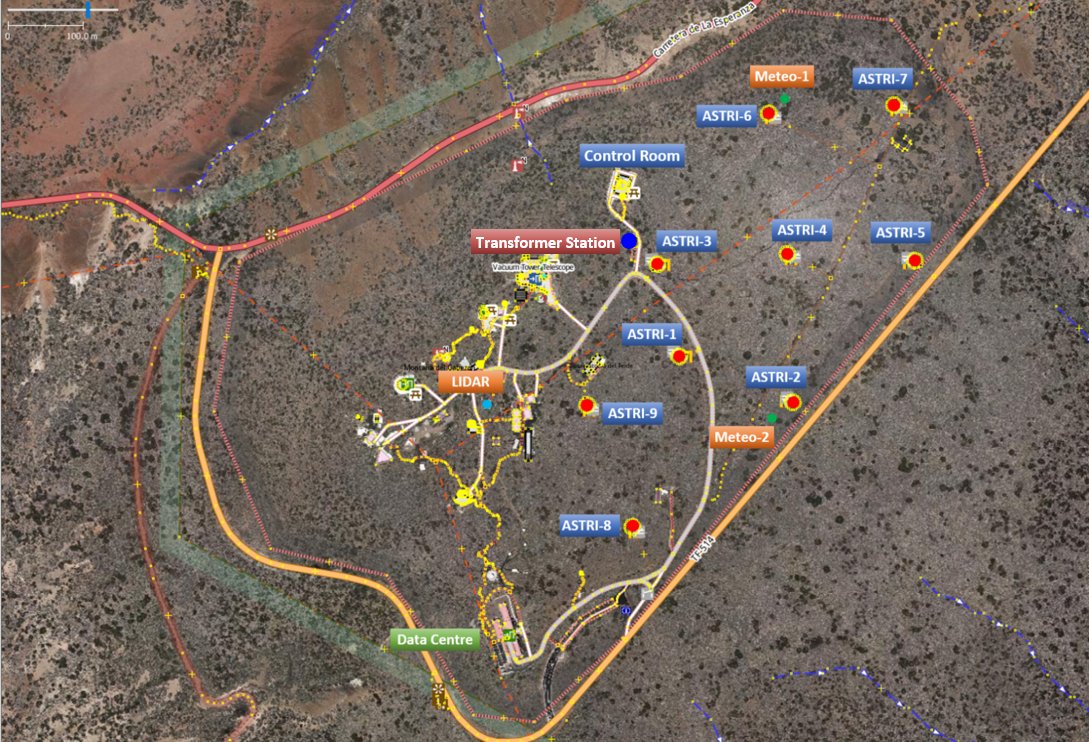}
	\caption{The final layout for the ASTRI mini-array. The Figure shows the final positions for the 9 telescopes, the position of the two meteorological towers, the position of the LIDAR, the position of the transformer station and that of the local control room and of the data centre. (Image obtained with Open Street Map editor)}
	\label{FIG:Layout}
\end{figure*}

The final layout of the ASTRI Mini-Array at the Teide Observatory is shown in Figure \ref{FIG:Layout}. The area where the telescopes will be installed will cover a rectangular strip of approximately 650 meters in length and 270 meters in width corresponding to a surface of about 17 hectares.

\begin{table}[width=.9\linewidth,cols=4,pos=h]
\caption{Geographical coordinates of the telescopes of the ASTRI Mini-Array}
\begin{tabular*}{\tblwidth}{@{} CCCC@{} }
\toprule
Telescope & Longitude W & Latitude N & Altitude \\
 & [$\degmark{}$ $\arcmin{}$ $\arcsec{}$] & [$\degmark{}$ $\arcmin{}$ $\arcsec{}$] & [m]\\
\midrule
ASTRI-1 & 28 18 03.69 & 16 30 28.69 & 2359 \\
ASTRI-2 & 28 18 02.43 & 16 30 23.78 & 2348 \\
ASTRI-3 & 28 18 08.52 & 16 30 29.82 & 2364 \\
ASTRI-4 & 28 18 08.31 & 16 30 23.90 & 2356 \\
ASTRI-5 & 28 18 08.73 & 16 30 17.63 & 2358 \\
ASTRI-6 & 28 18 14.92 & 16 30 24.88 & 2351 \\
ASTRI-7 & 28 18 15.56 & 16 30 18.56 & 2342 \\
ASTRI-8 & 28 17 57.45 & 16 30 31.34 & 2376 \\
ASTRI-9 & 28 18 02.75 & 16 30 33.98 & 2359 \\
\bottomrule
\end{tabular*}
\label{tab:GeoCoord}
\end{table}

The choice of the layout of the ASTRI Mini-Array was mainly driven by scientific performance. However, the characteristics of the area, the possibility to exploit available site infrastructures, the safety of operations and the impact on the activities of Teide observatory were also taken into consideration carefully. 

The position of the telescopes and their spacing has been analysed using Monte Carlo simulations. The simulations show that a symmetric layout with a spacing of about 250 meters yields the better compromise between a very good performance at high energies (>10 TeV) and a good sensitivity at lower ones (see for example, \cite{2016JPhCS.718e2008D}). Then the final layout has been chosen as a compromise among the necessity to preserve symmetry and optimum spacing, the shape of the available area and the possible constraints on the position of telescopes. The layout is asymmetrical being more elongated in the North-East/South-West direction with a median spacing among neighbouring telescope of about 200 m. The variation in performance due to the asymmetry of the array are expected to be relatively low in the whole energy range at low zenith angles ZA $\lesssim$ 30 degrees) of observation. In the case of higher ZA, dedicated Monte Carlo simulations are being produced in order to evaluate the effect.

The second driver to the layout was to take advantage of infrastructures (road, trenches, etc) already available at the site in order to reduce construction work and cost but also to minimize the impact on the ecosystem of the area where the Teide observatory is situated. In fact, the ASTRI Mini-Array will be placed in an area named ``Forest Crown'', near the national park ``Las Cañadas del Teide''. The area is of outstanding importance in natural landscape and vegetation. All efforts will be devoted to minimizing the impact on the environment both in the construction and in the operation of the telescopes.

Finally, the impact on the activities of the other scientific facilities at or nearby the Teide observatory will be considered in the development of the site first during construction activities and then in the operation phase. 

Table \ref{tab:GeoCoord} lists the geographical coordinates (latitude, longitude and elevation) of the nine telescopes of the ASTRI Mini-Array.

\section{The ASTRI Mini-Array}
\subsection{System architecture}
\begin{figure*}
	\centering
		\includegraphics[scale=1.3]{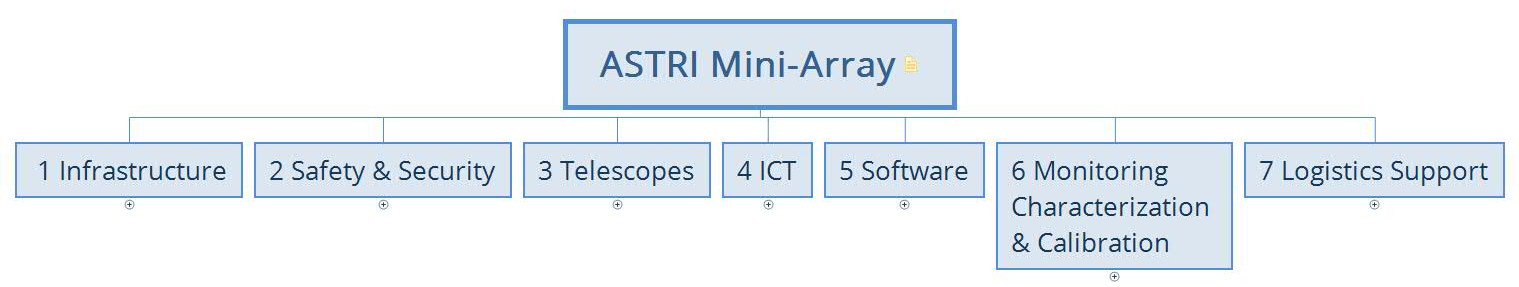}
	\caption{The ASTRI Mini-Array system decomposition showing the main subsystems.}
	\label{FIG:PBS}
\end{figure*}

\begin{figure*}
	\centering
		\includegraphics[scale=1.1]{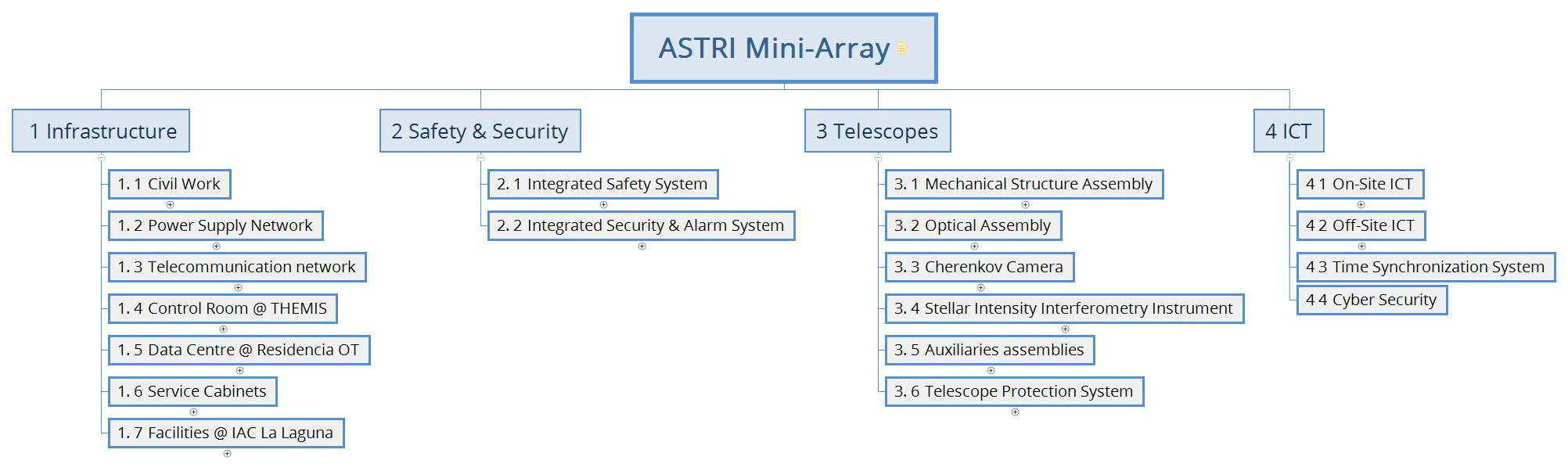}
	\caption{The ASTRI Mini-Array system decomposition down to the third level of the hierarchy. For the sake of clarity only 4 subsystems are displayed. (see also fig. \ref{FIG:PBS-2})}
	\label{FIG:PBS-1}
\end{figure*}

\begin{figure*}
	\centering
		\includegraphics[scale=0.3]{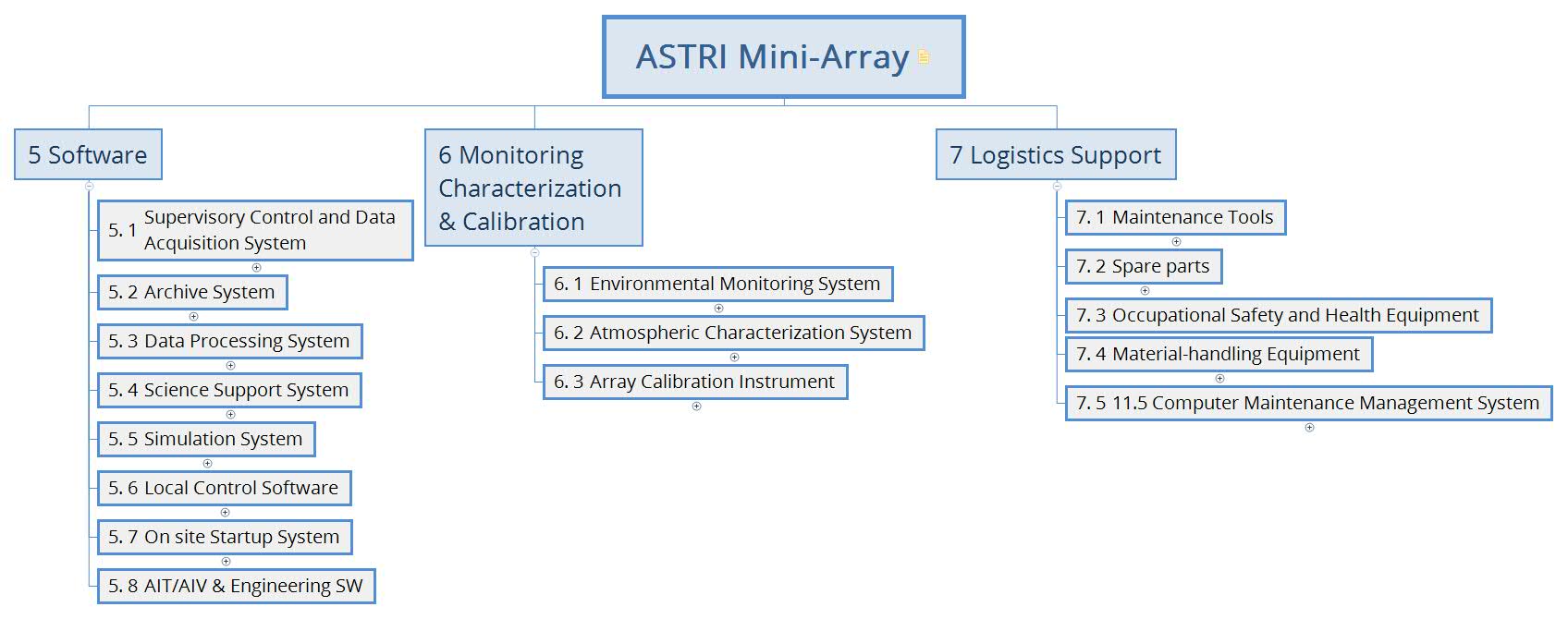}
		\caption{The ASTRI Mini-Array system decomposition down to the third level of the hierarchy. For the sake of clarity only 3 subsystems are displayed. (see also fig. \ref{FIG:PBS-1})}
	\label{FIG:PBS-2}
\end{figure*}

\begin{figure*}
	\centering
		\includegraphics[scale=0.45]{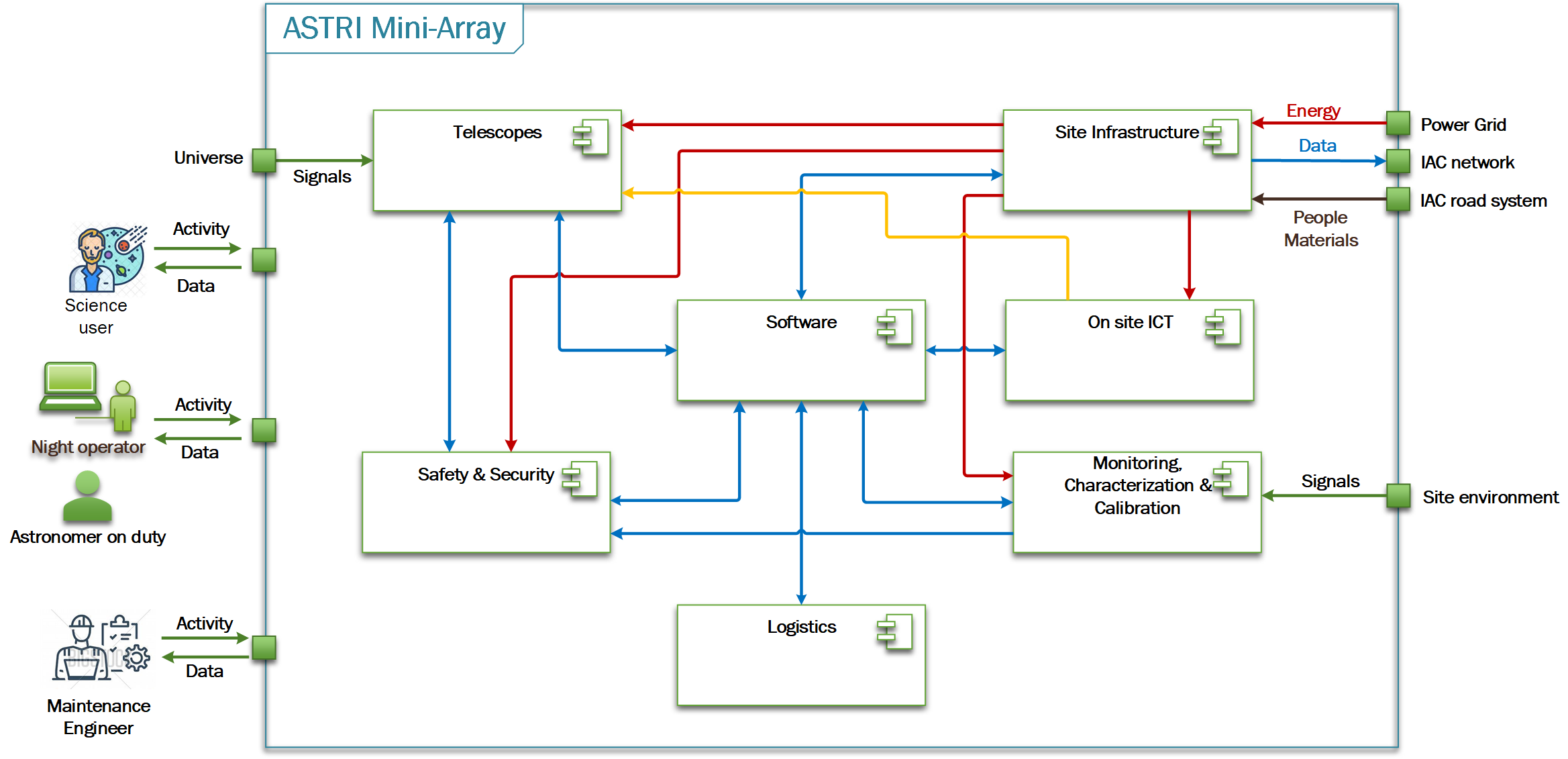}
	\caption{The ASTRI Mini-Array context diagram showing internal and external interfaces. Red line are for power connections, blue lines for commands and data, the yellow line is the time distribution system.}
	\label{FIG:context}
\end{figure*}

\begin{figure*}
	\centering
		\includegraphics[scale=1.5]{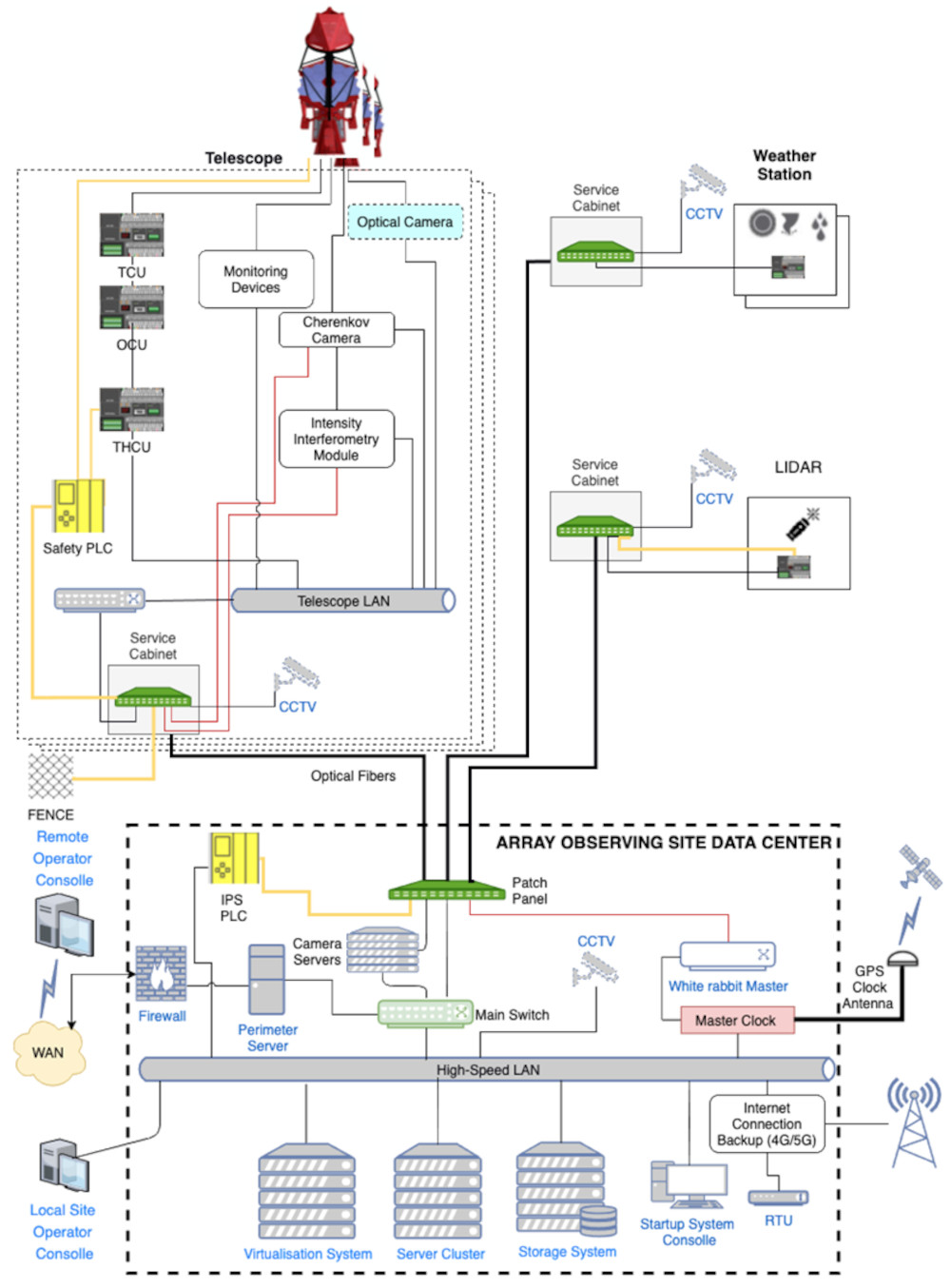}
	\caption{The ASTRI Mini-Array physical architecture. The black lines represent standard network cables or optical fibers; the yellow lines represent safety certified cables or optical fibers; the red lines represent white-rabbit certified cables or optical fibers.}
	\label{FIG:physical}
\end{figure*}

The architecture of the ASTRI Mini-Array is shown in Figures \ref{FIG:PBS}, \ref{FIG:PBS-1}, \ref{FIG:PBS-2}, \ref{FIG:context}, and \ref{FIG:physical}. Figure \ref{FIG:PBS} is the system decomposition down to the second level of the system hierarchy while Figures \ref{FIG:PBS-1} and \ref{FIG:PBS-2} go down to the third level.

The main subsystems making the ASTRI Mini-Array are:
\begin{enumerate}
    \item \textbf{Infrastructure}: composed by all those parts needed to make the observational site suitable to host the telescopes of the ASTRI Mini-Array.
    \item \textbf{Safety and Security System}: is an independent system for the protection of people and site assets.
    \item \textbf{Telescopes}: include the hardware used to collect and image Cherenkov light from air showers and stellar light for intensity interferometry measurements and the auxiliary assemblies needed to support this function. 
    \item \textbf{Information and Communication Technology (ICT)}: includes all computing/storage hardware, the overall networking infrastructure and all system services necessary on site and off site to control and monitor the array and to archive and analyse the scientific and engineering data.
    \item \textbf{Software}: the subsystem provides to the user a set of tools that will allow to perform all the operations from the preparation of an observing proposal to the execution of the observations, the analysis of the acquired data online and the retrieval of all the data products from the archive.
    \item \textbf{Monitoring, Characterization and Calibration Systems}: the set of devices that allows the environmental monitoring the atmospheric characterization and the array calibration.
    \item \textbf{Logistics}: includes all the hardware and software necessary for the preventive and corrective maintenance of the ASTRI Mini-Array.
\end{enumerate}
A brief description of the seven subsystems is given in the following paragraphs.

Figure \ref{FIG:context} instead is the general context diagram for the ASTRI Mini-Array system that displays the boundaries among the system and the external entities, identifies such entities that produces inputs and receive outputs. Interactions with the system are represented by arrows with a label identifying what is passed through the interface. So, for example, the Universe is one of the external entities that interacts with the system through the signals generated by photons emitted by the stars or in Cherenkov events. 

In the figure the ASTRI Mini-Array system is not represented as black box but its main subsystems are shown together with their internal interactions represented again by arrows this time with different colors depending on the type of quantity being provided or received. So, the yellow line connecting the on site ICT to the telescopes, represents the reference time with which all the events generated by the telescopes are tagged. Red lines represents the power distribution system through which the site infrastructure supplies all the ASTRI Mini-Array subsystems. Finally, the blue line represents the communication network through which the software subsystem sends commands to and receives data from all the other subsystems.

Finally, Figure \ref{FIG:physical} displays the physical architecture of the ASTRI Mini-Array. The focus of the figure is on the communication network. Three main networks are shown: the safety network, the time distribution system network and the generic network where data and commands are exchanged. The figure shows the connections between the on site ICT (named Data center in the figure) and the peripheral systems (telescopes, auxiliaries, operators) but also their internal connections.

\subsection{Site Infrastructure}
Figure \ref{FIG:PBS-1} shows the main elements that makes the site infrastructure subsystem:
\paragraph{Civil Work} 
The civil infrastructure deals with the design, analysis, management and construction of everything needed to support the deployment and the operation of the ASTRI Mini-Array, including foundations for telescopes and auxiliary instrumentation, roads connecting the telescopes, trenches for power and communication network. 
\paragraph{Power Supply Network} 
The power supply network includes the design, procurement and installation of all the hardware that, from the power transforming station at the Teide Observatory, is needed to supply electric power to all the mini-array subsystems. In the network are also included power backup systems (UPS and emergency power generator).  
\paragraph{Telecommunication Network} 
The communication network includes the design, procurement and installation of all the hardware necessary to the communication from the data centre to the telescopes and auxiliary instrumentation and from the data centre to the outside world.
\paragraph{Control Room at the Themis solar telescope}
This includes the infrastructure (power and data) necessary for the local control room hosted at the Themis solar telescope. 
\paragraph{Data Centre at the Residencia of the Teide Observatory}
This includes the infrastructure (power and data) necessary to host the onsite data centre, that necessary for safety and security and the air conditioning system.
\paragraph{Service Cabinets}
Service cabinets are the interfaces of the Telescope to the external world. In particular, they contain the power and communication networks interfaces. They also include part of the safety and security system (rain and humidity sensors, CCTV cameras).
\paragraph{Facilities at IAC in La Laguna}
Apart from the site infrastructure at the Teide Observatory, in La Laguna at the IAC the ASTRI project will need an office space to host the ASTRI personnel in charge of the coordination and the remote Mini-array Control room. Also, a warehouse will be needed to store spare parts and handling tools.

\subsubsection{The telescope area}
\begin{figure*}
	\centering
		\includegraphics[scale=0.6]{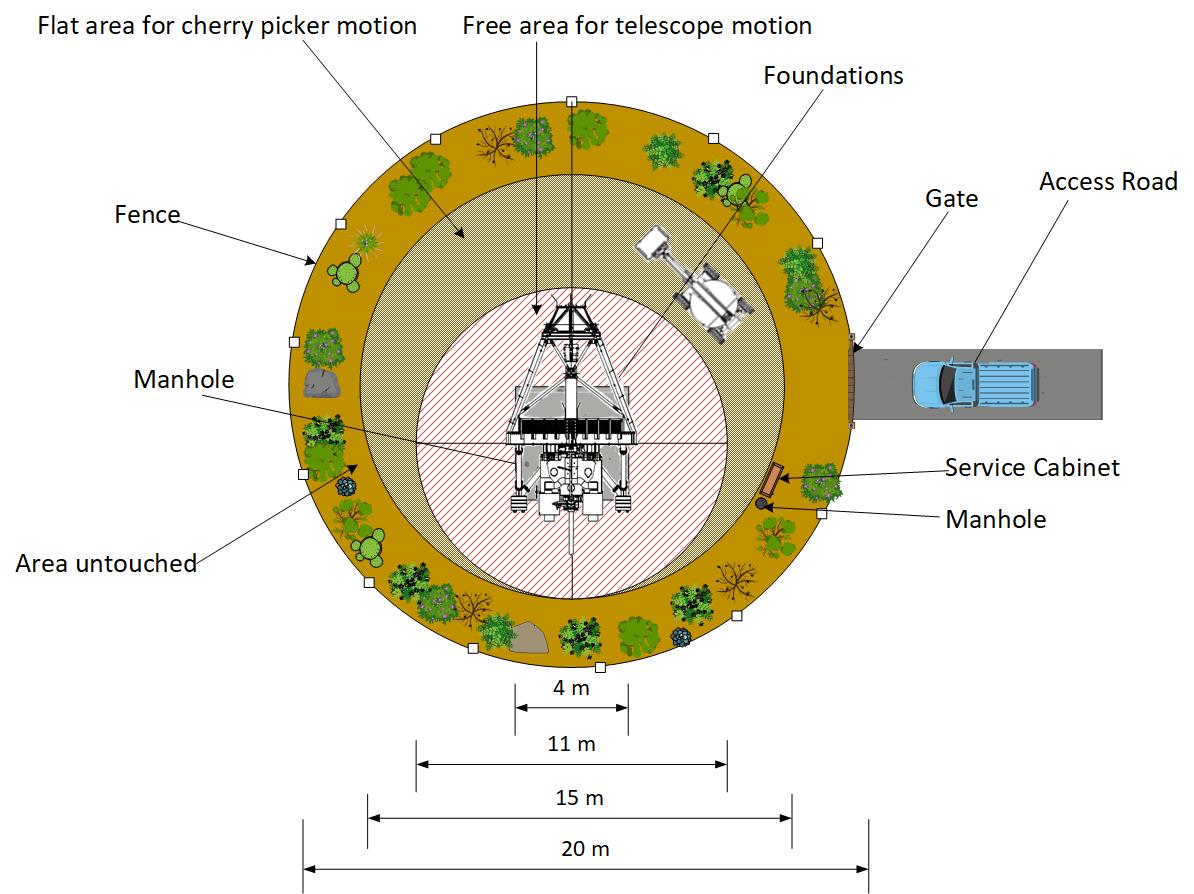}
	\caption{Proposed layout for the area around each telescope. A description of the area and its dimensions are given. A cherry picker and pick-up truck are also shown.}
	\label{FIG:TelArea}
\end{figure*}

The design of the area around each telescope is shown in Figure \ref{FIG:TelArea}. The telescope area includes:
\begin{enumerate}
    \item Telescope
    \item Foundations of the telescope 
    \item A service cabinet
    \item A security fence with an access gate
\end{enumerate}
Four zones can be identified inside the fence that delimits the telescope’s area:
\begin{enumerate}
    \item Foundations zone
    \item Telescope motion zone
    \item Cherry picker zone
    \item Untouched zone
\end{enumerate}
The foundations zone is the area occupied by the foundations of the telescopes and has a diameter of 4 meters.
The telescope motion free zone is a circular area centred on the telescope that has to be flat and free from any obstacle to allow the telescope to be moved undisturbed. This area shall have a diameter not less than 11 meters. 
The cherry picker zone is an area around the telescope of 15 meters of diameter that shall allow the use of a cherry picker for maintenance operations. The foreseen dimensions for the cherry picker will be 1.8 meters wide and 5 meters long when in parking position. The area is not centred on the telescope position because maintenance activity will happen mainly at the front of the telescope. The area shall be cleared and flat to allow safe motion of the cherry picker. On the border of this area a service cabinet will be placed. The service cabinet will be the interface between the cabinet and the external world with regard to power and communication.  
The untouched zone will be the area with an external diameter of 20 meters and internal diameter of 15 meters. This area does not need to be cleared or flat, so vegetation will be untouched. The untouched zone delimits the telescope area and is surrounded by a fence. The fence is needed for security and safety reasons and the access will happen only through a gate connected to the interlock safety system of the array. This zone is shown as circular and concentric to the cherry picker zone, but its actual shape will be adapted to the terrain conditions.

\begin{figure*}
	\centering
		\includegraphics[scale=1.9]{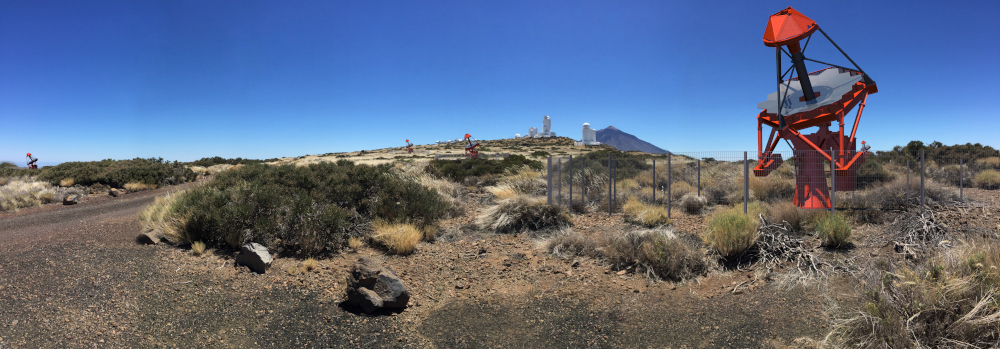}
	\caption{Artistic view of ASTRI Mini-Array Telescopes. From left to right: ASTRI-2, ASTRI-1, ASTRI-4, ASTRI-7 in the foreground surrounded by the fence, and ASTRI-6 in the background behind ASTRI-7. The picture shows an unrealistic scenario with telescopes pointing at the Sun. Credits: M. Leone.}
	\label{FIG:ArtView}
\end{figure*}

Figure \ref{FIG:ArtView} shows an artistic view of what part of the ASTRI Mini-Array will look like after construction.

\subsection{Safety and Security System}
The Safety and Security system (fig. \ref{FIG:PBS-1}) shall integrate all information available in real-time from any place on the site providing an interface to monitor the health status of all Site Elements. The system must implement safety functions to remotely stop the operations of all ASTRI Mini-Array subsystem in case of danger for people or things. The Safety and Security system is made by two subsystems:
\paragraph{Integrated Safety System} 
The integrated safety system includes the design and implementation of all the hardware and software needed to prevent unintended accidents to people or elements of the ASTRI Mini-Array during operations or maintenance activities. Typical elements of such a system are interlocks, emergency stops, fire alarms. This system will be completely independent of any other system present at the observing site apart from power. 
\paragraph{Integrated Security and Alarm System}
This system is intended to protect property and environment of the ASTRI Mini-Array assets from external “threats”. It will be composed of an access control system to prevent unauthorized people from accessing telescope areas or data center and closed-circuit television cameras (CCTV).

\subsection{The telescopes}
\label{sec:Telescope}
\begin{figure}
	\centering
		\includegraphics[scale=0.6]{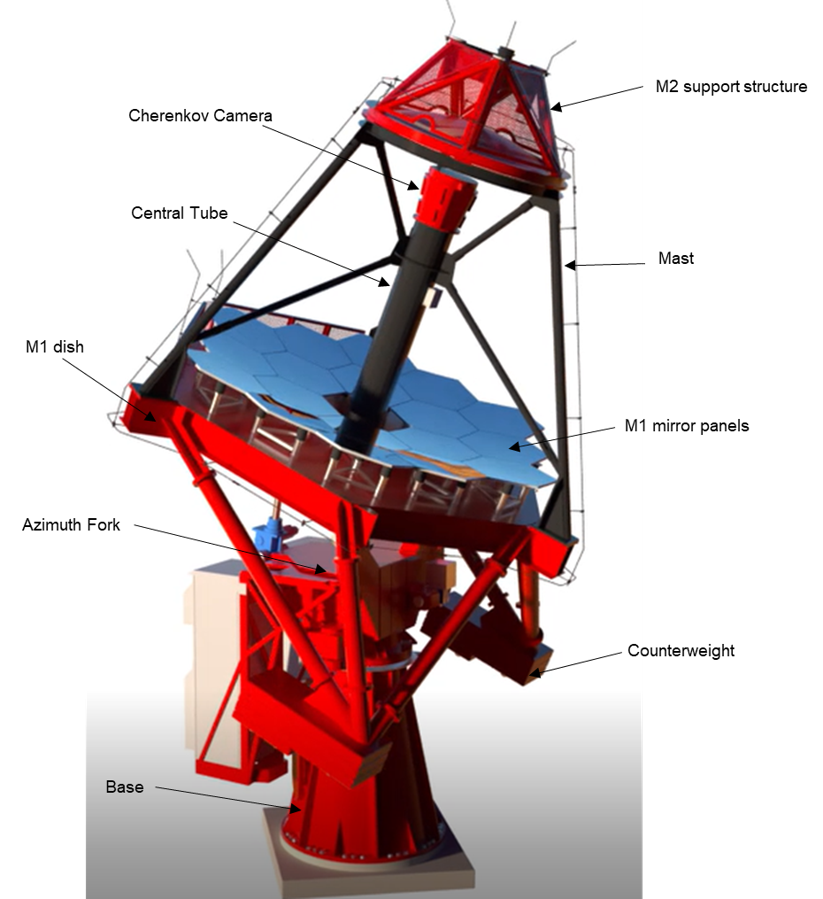}
	\caption{3D model of a telescope of the ASTRI mini-array. Some of the opto-mechanical components are explicitly listed.}
	\label{FIG:Telescope}
\end{figure}

The Telescope subsystem includes mainly the hardware assemblies to collect and image Cherenkov light from air showers and star light for the stellar intensity interferometry instrument and the auxiliary assemblies needed to support these functions. For each assembly a local control software will also be developed to manage the functionality of the assembly itself.  
Figure \ref{FIG:Telescope} shows a 3D model of one of the ASTRI telescopes to be deployed at the Teide Observatory site.
The main subsystems of the ASTRI telescope are:
\begin{enumerate}
    \item Mechanical Structure 
    \item Optics
    \item Cherenkov Camera
    \item Stellar Intensity Interferometry Instrument (SI$^3$)
    \item Auxiliary assemblies
\end{enumerate}

\paragraph{Mechanical Structure} 
The ASTRI telescope adopts an alt-azimuthal design in which the azimuth axis will permit a rotation range of ±270 $\degmark{}$. The primary mirror dish, which supports the primary mirror, is mounted on an azimuth fork, which allows rotation around the elevation axis from 0 to +91 $\degmark{}$. The mast structure, that supports the secondary mirror and the camera, is placed on the primary mirror dish. 
The current design of the ASTRI electro-mechanical structure is an evolution of the ASTRI-Horn prototype telescope. Exploiting the lessons learned during the operation of the prototype the electro-mechanical structure has been optimized in terms of mass, functionality and maintainability. In particular the mass has been reduced by 30\% going from 25 tons down to 17.5 tons while keeping the stiffness of the structure. The primary active control system is not permanently mounted on the telescope but has been ``degraded'' to an Assembly, Integration, and Validation (AIV) tool and to a maintenance tool. Modifications to the secondary mirror support structure and to the primary mirror dish structure brought to easier maintenance procedures and less vignetting. A detailed description of the improved electro-mechanical structure can be found in \citep{2018SPIE10700E..5WM}.

\begin{figure}
	\centering
		\includegraphics[scale=0.21]{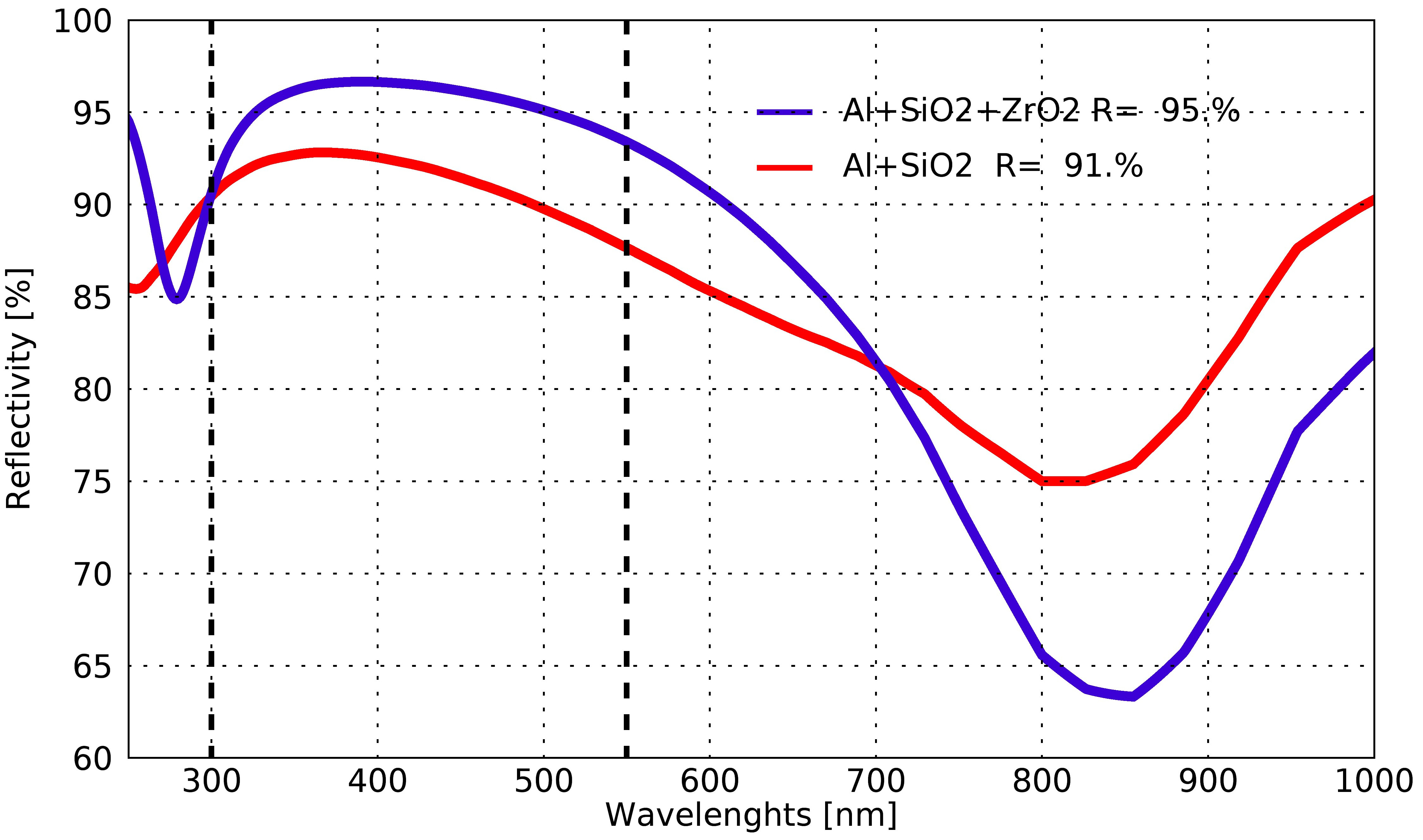}
	\caption{Comparison between the Al+SiO$_2$ coating used in the ASTRI-Horn prototype telescope and the one (Al+SiO$_2$+ZrO$_2$) used for the mirrors of the ASTRI Mini-Array telescopes. The vertical dashed lines mark the region of interest of Cherenkov astronomy.}
	\label{FIG:Coating}
\end{figure}

\paragraph{Optics} 
The optical design is based on a modified (\citep{2007APh....28...10V}, see also \citep{2017SPIE10399E..03S}) Schwarzschild-Couder configuration. This configuration allows better correction of aberrations at large incident angles even for small focal ratios and hence facilitates the construction of compact telescopes. This optical system is an attractive solution since it enables good angular resolution across the entire field of view and allows reducing the focal length and therefore the physical pixel and overall camera size. The primary mirror (M1) is segmented while the secondary (M2) is monolithic. The primary is composed by a set of 18 hexagonal-shaped panels. The profiles of both mirrors are aspheric with substantial deviations from the main spherical component.
The primary mirror has a diameter of 4.3 m while the secondary mirror diameter is 1.8 m. The primary-to-secondary distance is 3 m and the secondary to camera distance 0.52 m. 
This optical setup delivers a plate scale of 37.5 mm/degree, an equivalent focal length of 2150 mm and an effective area of about 5 m$^2$. The optical design of the Mini-Array telescope has remained unchanged respect to ASTRI-Horn but the coating of the mirrors was improved using a combination of Al, SiO$_2$and ZrO$_2$. Figure \ref{FIG:Coating} clearly shows that the new coating performs better in the wavelength region of interest of Cherenkov astronomy, see \cite{2022JATIS...8a4005L} for a description of the characterization of the ASTRI mirrors.

\paragraph{Cherenkov Camera} 
The small plate scale resulting from the optical design leads to a Cherenkov camera of compact dimensions. This, in turn, allows the use of silicon photomultipliers (SiPM) as focal plane detectors. The SiPM detectors chosen for the ASTRI Mini-Array Cherenkov cameras are those of LVR3 series produced by Hamamatsu photonics. They are uncoated and have linear dimensions of 7x7 mm. To cover the corrected field of view of the telescope 2368 pixels are needed. The pixels are grouped in matrices of 8x8 pixels. 37 matrices are arranged to adapt to the curved focal plane of the telescope. 
The characteristics of the optical system coupled with the physical characteristics of the SiPM detectors yields an angular pixel size of 0.19 degree and a field of view of 10.5 $\degmark{}$. Furthermore, more than 80\% of the light emitted by a point source is collected within the dimensions of a pixel over the full field of view of the telescope.

The ASTRI camera read-out electronics ensures a high efficiency detection of a Cherenkov event, with over 1000 events per second, with a very high dynamical range (from 1 to 1500 photoelectrons per pixel) clearing the way to the possibility to observe even in moderate moon illumination conditions. At the hearth of the electronics is the CITIROC-1A ASIC that implements a customized signal shaper and peak detector to acquire the SiPM pulses. The camera trigger is a topological one, activated when a given number of contiguous pixels presents a signal above a threshold equivalent to a given number of photo-electrons.  

The last innovation introduced in the ASTRI camera is the use of an interferential filter as front window (\citep{2018NIMPA.908..117R} and \citep{2018SPIE10702E..37C}) that allows to reduce the contribution from the night sky background at wavelengths greater than 550 nm where the sensitivity of SiPM detector is still high. 

\begin{figure}
	\centering
		\includegraphics[scale=0.29]{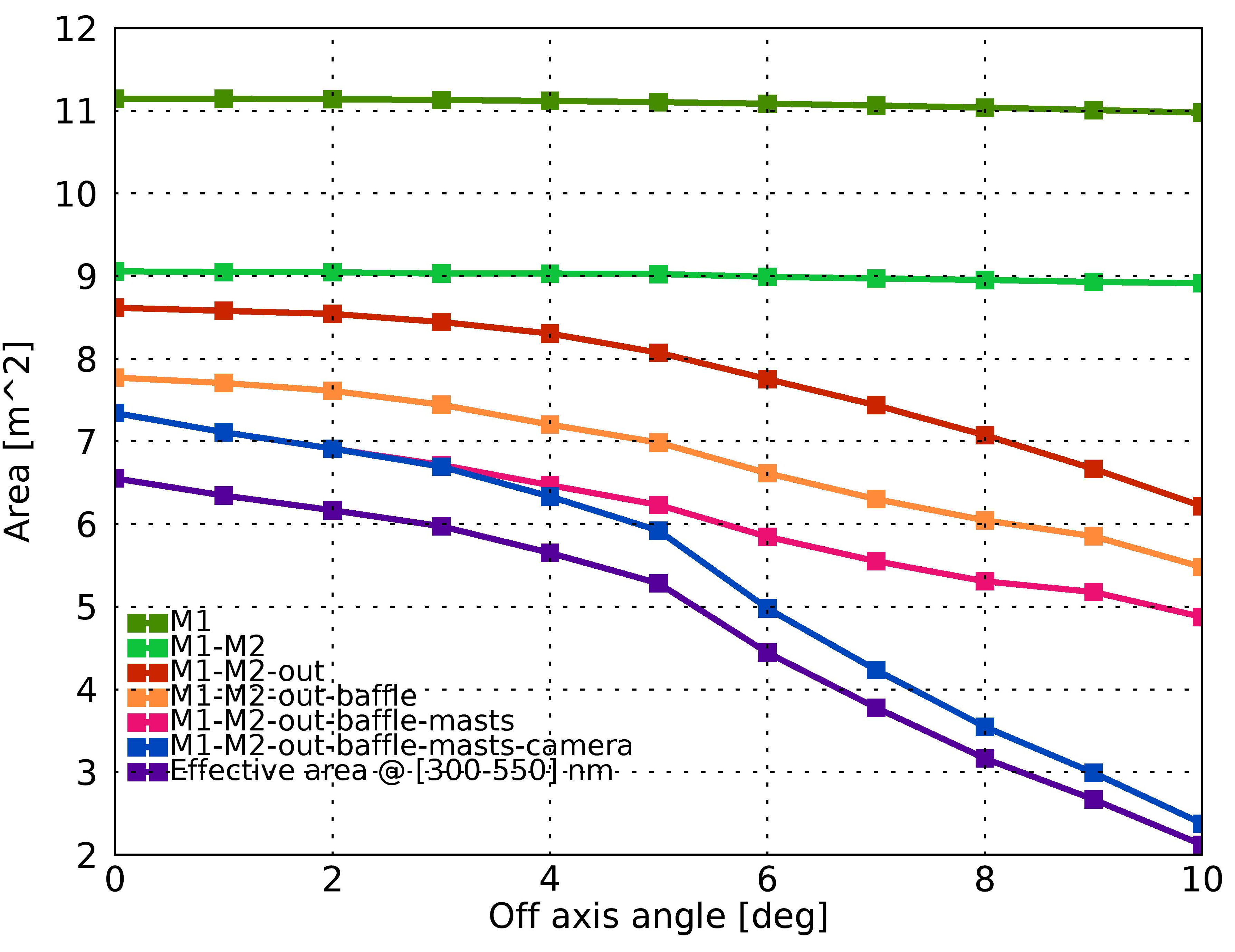}
	\caption{The effective area of the ASTRI telescope as a function of the angular position on the field of view. The different curves shows the reduction due to the vignetting of the various parts of the telescope. The magenta curve includes also the contribution of the camera filter.}
	\label{FIG:Vignetting}
\end{figure}

\begin{figure}
	\centering
		\includegraphics[scale=0.29]{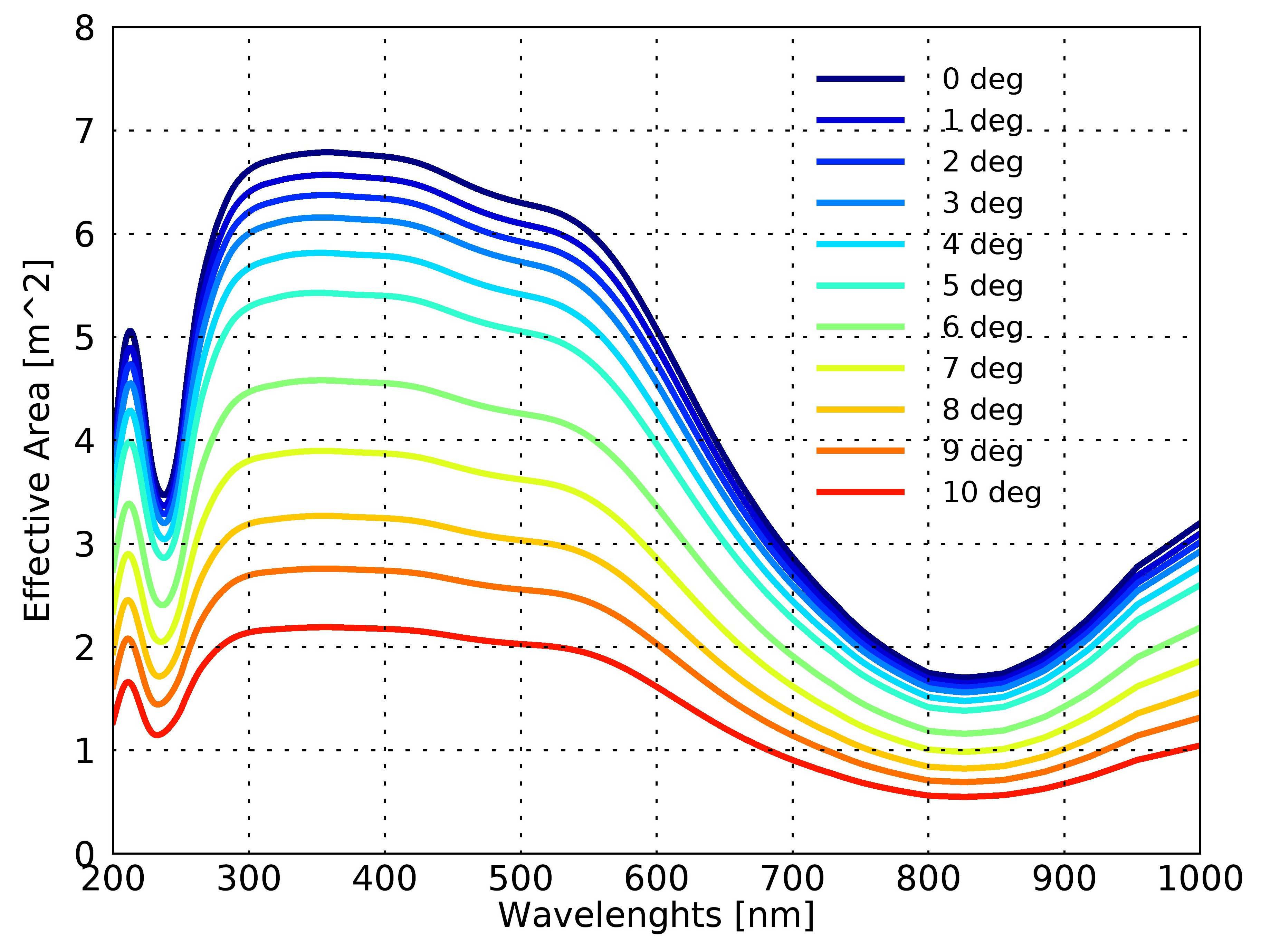}
	\caption{The effective area of the ASTRI telescope as a function of wavelength. The different curves refer to different angular positions on the field of view.}
	\label{FIG:Effective}
\end{figure}

Figures \ref{FIG:Vignetting} and \ref{FIG:Effective} show the effective area of the telescope as a function of angular position on the field of view and wavelength respectively. In particular, Figure \ref{FIG:Vignetting} shows that, taking into account all the effects of the vignetting of the various parts of the telescope and the reduction due to the Cherenkov camera interferential filter, the effective area is always above 5 m$^2$ throughout the entire field of view of 10.5$\degmark{}$, decreasing of 25\% when going from the center to the edge of the field of view. Figure \ref{FIG:Effective} shows the effect of the filter in reducing the signal at wavelengths larger than 550 nm.

\paragraph{Stellar Intensity Interferometry Instrument (SI$^3$)}
The Stellar Intensity Interferometry Instrument is a dedicated optical photon detection module for performing intensity interferometry observations with the ASTRI Mini-Array telescopes. 

\begin{figure}
	\centering
		\includegraphics[scale=0.8,angle=0]{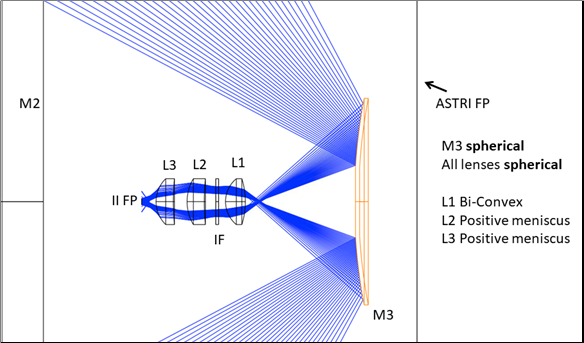}
	\caption{Close-up view of the SI$^3$ catadioptric module deployed above the telescope Cherenkov focal plane (ASTRI FP). The beam is reflected off M3 and collimated and refocused by an objective lens group onto the intensity interferometry focal plane (II FP). M2 is the telescope secondary mirror}
	\label{FIG:SI3}
\end{figure}

The optical layout of the SI$^3$ instrument is shown in \ref{FIG:SI3}. A convex spherical mirror (M3 in the figure), placed on top of the Cherenkov camera, is used to reflect back the light from M2 mirror in an optical module, made by a series of spherical lenses, that collimates the light on a narrow band filter and then focuses it onto the detector. The detector is a 2x2 array of SIPM pixels with 3x3 mm linear dimensions produced by Hamamatsu Photonics. The readout electronics is based on the MUSIC ASIC able to sustain a rate of 100MHz per channel (pixel). The instrument will be placed in position in front of the ASTRI Cherenkov camera using a swing mechanical arm, 

\paragraph{Auxiliary Assemblies}
The auxiliary assemblies are those items that support the main functions of the telescope during operations and maintenance. In particular:
\begin{itemize}
    \item \textit{Pointing monitoring camera}: a CCD camera placed on the M2 support structure used to monitor pointing and tracking performances of the telescope.
    \item \textit{Mirror alignment system}: a removable set of actuators for M1 and an optical camera used to align the panel of M1.
    \item \textit{Telescope condition monitoring system}: a set of temperature sensors and accelerometers used for the predictive maintenance of the telescope.
\end{itemize}

\subsection{Information and Communication Technology}

\paragraph{On Site ICT}
The on-site Information and Communication Technology system that shall be installed at Teide includes all computing/storage hardware, the overall networking infrastructure (including cabling and switches) and all system services (operating system, networking services, name services, etc.) to control the array and monitor its health status, perform online observation quality analysis, store temporarily data at the site and guarantee internal and external network communications.

\begin{figure}
	\centering
		\includegraphics[scale=1.2]{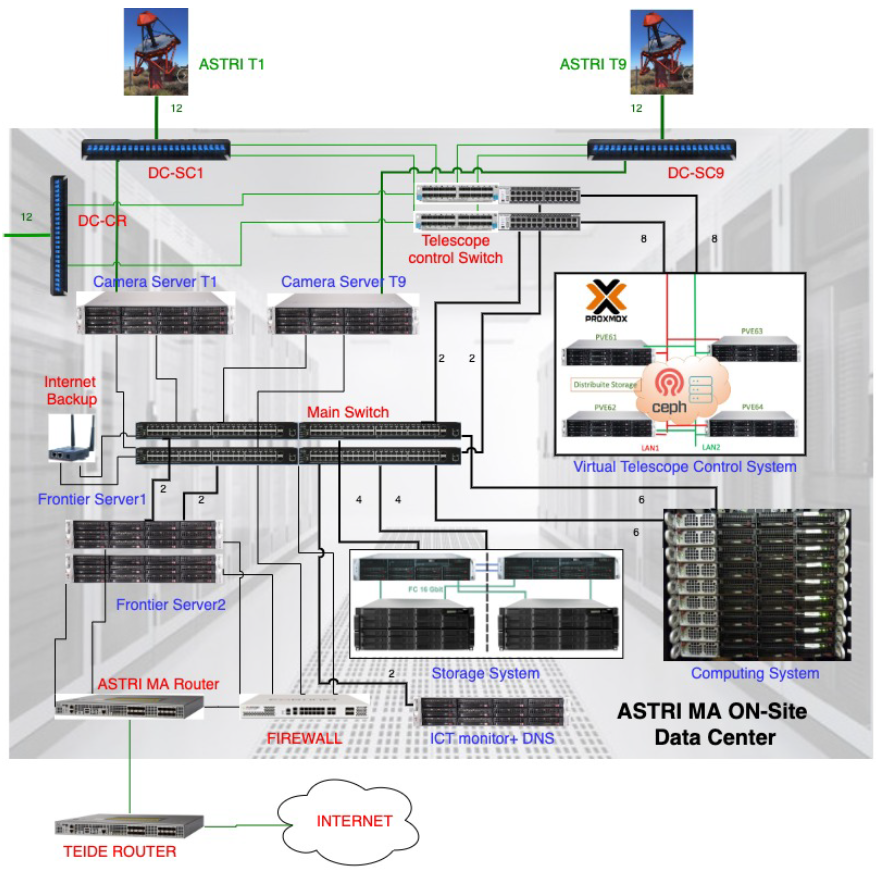}
	\caption{The physical architecture of the ASTRI Mini-Array on site ICT.}
	\label{FIG:ICT}
\end{figure}

The on-site Information and Communication Technology system architecture is shown in Figure \ref{FIG:ICT} with the following main elements:
\begin{itemize}
    \item \textit{Virtual Telescope Control System}: the system hosting the virtual machines that will be used for the telescopes control.
    \item \textit{Camera Servers}: are the physical servers, one for each telescope, for the Cherenkov camera and stellar intensity interferometry data acquisition.
    \item \textit{Computing System}: is the set of physical servers dedicated to the on-line analysis of scientific data for quality check and of monitoring data for the alarm management.
    \item \textit{Storage System}: is the collection point of the raw scientific data, of the monitoring and of the alarm data. It also the location from where all these data are accessible for remote transfer and for all on-site uses.
    \item \textit{Network System}: is the set of devices responsible for internal and external network connections. 
\end{itemize}

\paragraph{Off Site ICT}
The off-site Information and Communication Technology system that shall be installed at INAF Rome observatory includes computing/storage hardware, overall networking infrastructure and all system services.  
At the off-site ICT all the data produced by the mini-array will be archived and analysed. Also, it will host a gateway for science users of the mini-array.

\paragraph{Time Synchronization System}
The Time Synchronization System will be a Timing Distribution System designed to keep clocks synchronized to sub-ns precision for Cherenkov Event Timing purposes. This system will allow to time tag any event recorded by every Cherenkov camera of the array.
The system will be based on the white-rabbit technology and will be deployed at the mini-array observing site. 

\subsection{Software}
\begin{figure*}
	\centering
		\includegraphics[scale=0.5]{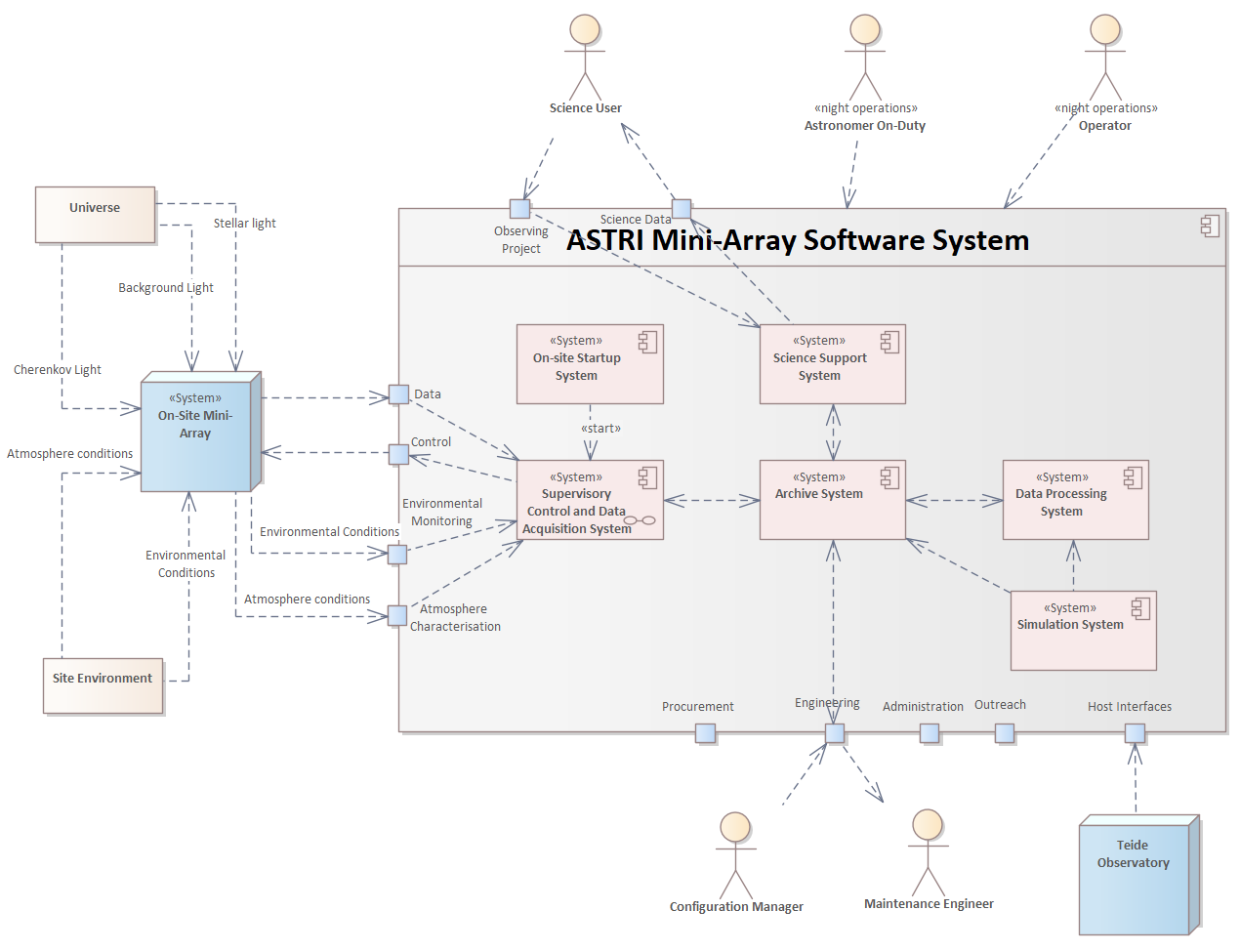}
	\caption{Context view of the ASTRI Mini-Array Software System. The main software subsystems are shown.}
	\label{FIG:Functional}
\end{figure*}

Figure \ref{FIG:Functional} shows the context diagram of the Mini-Array software system with the main software subsystems and their internal and external relationships.

The main software subsystem are the following (see also Figure \ref{FIG:PBS-2}):
\paragraph{Supervisory Control And Data Acquisition (SCADA) System}
The software system devoted to control all the operations carried out at the MA site, including the startup of the MA system. SCADA is a central control system which interfaces and communicate with all equipment and dedicated software installed On-Site.
\paragraph{Archive System}
The software service that provides storage and organization for all data, data products, and metadata generated for and by the Mini-Array, and defined by the Mini-Array Data Models.
\paragraph{Data Processing System}
The software system used to calibrate and reduce the data acquired. This software is also used to check the quality of the final data products.
\paragraph{Science Support System}
The software system which provides the main point of access for the exchange of science-related data and information with the ASTRI Science Users, and which supports the whole science-related workflow, from the Observing Project submission to the access to the archived high-level Mini-Array science data products and the corresponding Science Tools to support data analysis.
\paragraph{Simulations System}
The software system that runs Monte Carlo simulations to provide simulated data for the development of reconstruction algorithms and for the characterization of real observations.
\paragraph{Local Control Software}
Firmware and low-level software dedicated to the low-level hardware control operations.

The ASTRI Mini-Array software is envisioned to handle an observing cycle, i.e. the end-to-end control and data flow system. The observing cycle can be divided into the following main phases:
\begin{enumerate}
    \item Observation preparation;
    \item Observation execution;
    \item Data Processing;
    \item Dissemination.
\end{enumerate}

\begin{figure*}
	\centering
		\includegraphics[scale=1.5]{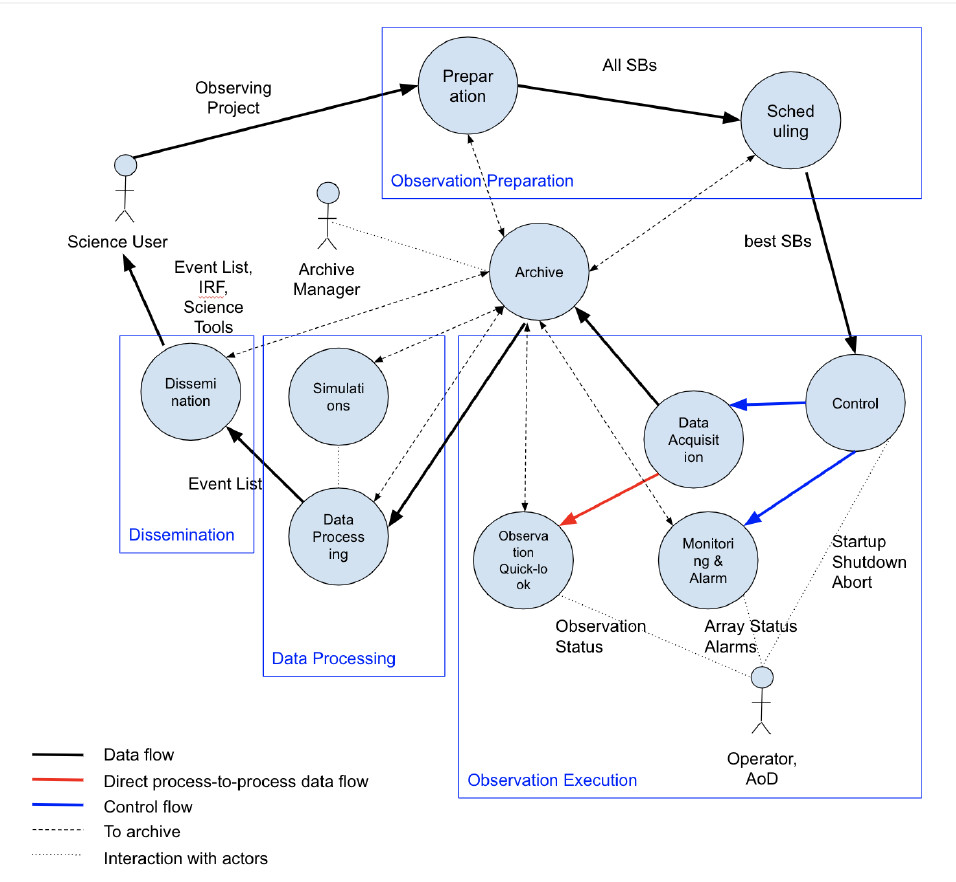}
	\caption{MA Software System data and information flow (schematic) with the four main phases. The outer solid black and red lines show the logical data flow, where the solid blue lines are control flow. Direct process-to-process communication is indicated with a red line. The dashed lines directed to/from the Archive indicate that a) all data are saved and can be retrieved from the Archive, and b) that the physical data flow may be handled by the Archive.}
	\label{FIG:Flow}
\end{figure*}

Figure \ref{FIG:Flow} shows how the data and information flow during an observing cycle. The central role of the archive is clear. 

\subsection{Monitoring, Characterization, and Calibration System}
\paragraph{Environmental Monitoring System}
The Environmental Monitoring System includes all the hardware necessary to monitor the atmospheric weather conditions at the observing site. In particular, two meteorological stations will be installed on telescopic towers. They will be solar powered and radio connected to the ASTRI Mini-Array network. Rain sensors and humidity sensors will be mounted on the service cabinet. 
\paragraph{Atmospheric Characterization System}
The Atmospheric Characterization System includes all the devices necessary to measure the Night Sky Background (UVSiPM and SQM) and the atmospheric extinction (LIDAR) in the observing direction. These data will be used to properly correct the Cherenkov measurements.
Specifically, the LIDAR will allow to measure:
\begin{itemize}
    \item \textbf{Atmosphere Transmission}: the value of the atmospheric transmission will be used during the reduction/analysis of the ASTRI-MA data. The atmospheric transmission will be obtained by LIDAR data, during regular data taking, pointing to a region of the sky close by the RA-Dec coordinates under investigation.
    \item \textbf{Atmosphere Extinction Profiles}: the atmosphere extinction profiles are used to define the atmospheric model needed for the Monte Carlo simulations. Daily profiles can be obtained from the analysis of LIDAR data acquired at sunset before the start of the Cherenkov observations, and at dawn after the end of observations. 
\end{itemize}

\paragraph{Array Calibration System}
The array calibration system is made only by one device, the illuminator. This portable ground-based device is designed to uniformly illuminate, from a certain distance, the telescopes aperture with a pulsed or continuous reference photon flux, and it will be used to perform the absolute end-to-end calibration of each telescope of the ASTRI Mini-Array. The system allows us to measure the actual spectral and temporal response of each ASTRI Mini-Array telescope at any off-axis angle, by changing the telescope pointing with respect to the Illuminator. Results will allow to obtain, for groups of few pixels, the corrective factor for the flat fielding of sky images. See \cite{2016SPIE.9906E..3SS} for a more detailed description of the illuminator concept.

\subsection{Logistics Support}
The logistics support includes all the hardware and software necessary for the preventive and corrective maintenance of the ASTRI Mini-Array.
\paragraph{Maintenance Tools}
All the hardware tools necessary for the preventive and corrective maintenance of the mini-array elements. Among these tools are those specifically designed and manufactured for the mini array necessities (like the Camera Handling Tool) or those completely generic (as greases or wrenches).
\paragraph{Spare parts}
All the hardware kept in an inventory and used for the repair or replacement of failed units of the mini array. 
The type and quantity of each spare part is the result of the RAM analysis of the system. For each subsystem spare are kept only at the level of LRU as defined for that subsystem. For example, in the case of the Cherenkov camera, the LRU is the camera itself.
\paragraph{Occupational Safety and Health equipment}
All the equipment needed for personal safety and health (safety helmet, gloves, shoes, first aid kits, etc.).
\paragraph{Material Handling Equipment}
All equipment needed to handle any material necessary for the installation and maintenance of the ASTRI mini array subsystems.
\paragraph{Computer Maintenance Management System}
The Compute Maintenance Management System will be a commercial software package that maintains a computer database of information to organize the Mini Array maintenance operations.

\section{The implementation of the ASTRI Mini-Array}
\begin{figure*}
	\centering
		\includegraphics[scale=0.27]{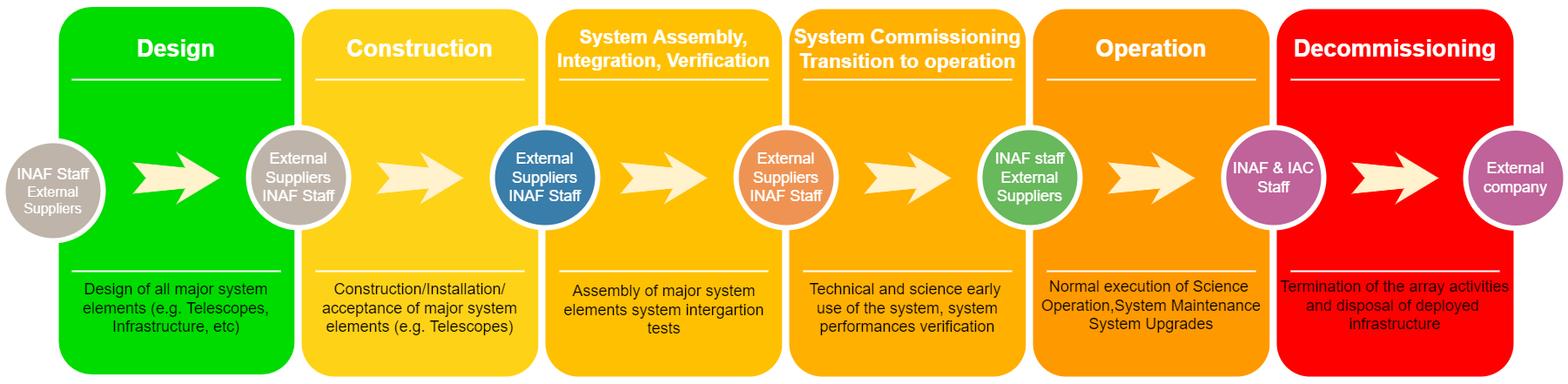}
	\caption{The ASTRI Mini-Array project phases from design to decommissioning}
	\label{FIG:Lifecycle}
\end{figure*}

Figure \ref{FIG:Lifecycle} shows the phases that will take the ASTRI Mini-Array project from design to decommissioning through construction and operation.

\subsection{Design Phase} 
During the design phase the concept of mini-array, its architecture and the top level requirements will be developed to produce a detailed design. Then requirements and design of the various subsystems will be produced. At the moment of the writing of this document several subsystems have already concluded the design phase (e.g. telescope, optics) while other are in it (infrastructure, software, Cherenkov camera).  
\subsection{Construction Phase}
During the construction all the elements making the ASTRI Mini-Array will be constructed, procured, manufactured or developed. 
Direct construction will have to do only with all the infrastructures necessary to make the observing site suitable to host and operate the ASTRI Mini-Array. The implementation of the site has been assigned through a tender that will include all the necessary activities.
All the major subsystems (e.g. mechanical structure, mirrors, cameras, ICT) will be procured through industrial tenders. 

\subsection{AIV Phase}
The AIV phase consists of the assembly, integration and verification of all ASTRI Mini-Array subsystems.

The prerequisite to the start of this phase is the completion of all infrastructure works. This includes all civil works, the power, and the telecommunications network. Safety and Security system should also be completed.

The first subsystem that will go through the AIV will be the onsite ICT, followed by auxiliary devices (meteo station, all sky camera etc). Software for onsite array control and data handling will be deployed and the telescopes will follow right after. 
The scheme of the AIV of the single telescope will be as follows:
\begin{enumerate}
    \item Assembly, Integration and Verification of the mechanical structure. The responsibility for this activity will be on the industrial contractor.
    \item Integration and Verification of optics. The responsibility of this activity will be of the INAF personnel that will be assisted in this task by the telescope industrial contractor. 
    \item Integration and Verification of the Cherenkov Camera. For this activity the camera industrial contractor and the ASTRI Team will share responsibility working together during the integration of each camera and the execution of the agreed operative test sequence necessary to verify the performances, functions and interfaces with the telescope.
    \item System tests on the array functionalities will start as soon as two telescopes will conclude their specific AIV activities.
\end{enumerate}

\subsection{From Commissioning to Operation}
The commissioning phase will allow the ASTRI team to verify the performance of the ASTRI Mini-Array from a technical and a scientific point of view. Deep in-system evaluation of all components and software will be carried out. Furthermore, during this phase all operation modes will be tested. 

The final activity of the commissioning phase will be the end-to-end verification of the ASTRI Mini-Array through the execution of the Science Verification program developed by the ASTRI science team.
\subsection{Operation (Scientific) Phase}
During this phase regular science operations, maintenance activities and data processing will take place. 
\subsection{Decommissioning Phase}
During this phase an external company will dismantle the telescopes and all the other ASTRI Mini-Array infrastructures and the site will be returned to its pristine state.

\subsection{Current timeline of the project}
With the exception of the SI$^3$ instrument, and the LIDAR all the other subsystems are in the construction phase with the AIV of the first telescope scheduled Summer 2022. The current master schedule foresees a first set of three telescopes completed by Summer 2023, while the full array will be ready for commissioning by beginning of 2025 with the operations starting four months later.

\section{Science operation plan}

\subsection{Operation Concept}

For the initial 3 year period the ASTRI Science team will develop a strategy to concentrate the ASTRI Mini-Array observational time on a limited number of programs with clearly identified objectives (see Paper II). This will produce an observation schedule that will span several months at a time.

No real time analysis of the data is foreseen but only a data quality check. Data analysis policy adopted will then be next day processing. 
No array trigger (stereo trigger) will be implemented at the site. Every Cherenkov event detected by any telescope will be time tagged and stored. Any search for Cherenkov events detected in coincidence by more than one telescope will be performed via software off-line at the Rome Data Center.
No subarray operation is foreseen apart from possible tests during the construction phase.
Night Science operation will be controlled from a remote control room so no people are required to be present at the observational site during the night. 

\subsection{Operation modes}
The ASTRI Mini-Array will implement the following type of operation modes:
\begin{itemize}
    \item \textbf{Normal observation mode}: to observe the targets as defined by the Science Operation Plan. Usually science observations require dark time, within nautical twilight, although it is possible to operate also during moderate moonlight conditions. Calibration activities are included in the normal operation mode.
    \item \textbf{ToO Mode}: the science operation plan will identify some astrophysical targets (either a specific object or a class of objects) that, giving raises to transient phenomena, will require a response from the night operator and a change in the night schedule. This means that no dedicated automatic software procedure to react to these transient phenomena is foreseen.  Depending on the type of transient object the reaction time will vary from 1 hour to 1 day.
    \item \textbf{Coordinated Mode}: possible synergies with the current and future VHE arrays (HAWC, MAGIC, LST, CTAO, and LHAASO) in the northern hemisphere are foreseen. This means that simultaneous observations, scheduled in advance, with some of these arrays, specifically MAGIC and LST, will be possible. Other synergies with optical telescopes operational at the Canary Islands (e.g. TNG) will also be possible.
    \item \textbf{Maintenance mode}: this mode deals with all activities necessary for the maintenance of the telescopes, the on-line control software, the monitoring, characterization and calibration devices, and the infrastructures (e.g. network, data center, etc.). This is the only daytime operation mode.
\end{itemize}

\subsection{Observing plan}

Table \ref{tab:AAOT} shows the Average Annual Observation Time (AAOT) for the Observatorio del Teide. The AAOT contains several factors \citep{2014SPIE.9145E..0KS}, the number of nights without moon corrected for the fraction of nights with clear sky and for the fractional loss for bad weather or ``Calima'' a dust wind originating in the Saharan Air Layer. Taking into account also the time loss for technical reasons (10\%) we end up with 1000 hours per year of available time for Cherenkov observations in dark conditions. This corresponds to a duty cicle of about 11.5 \% that is quite typical for IACT systems (see, for example, \citep{2017APh....94...29A}).
Stellar Intensity Interferometry does not need dark time so it will be performed around full moon (3 to 4 days per month).  

\begin{table}
\caption{Average Annual Observation Time for the Observatorio del Teide.}
    \centering
    \def\arraystretch{1.5}
    \begin{tabular}{|l|r|}
    \hline
    Moonless Night Hours                              & 1565 h \\
    Fraction of clear nights (cloud coverage $<$20\%) & 0.79 \\
    Fractional loss due to bad weather                & 0.04 \\
    Fractional loss due to "Calima''                  & 0.07 \\
    \hline
    Average Annual Observation Time                   & 1104 h\\
    \hline
    \end{tabular}
    \label{tab:AAOT}
\end{table}

Table \ref{tab:Pillars} is adapted from Tables 3 and 5 from Paper II and gives the tentative lists of targets for the first years of operation of the ASTRI Mini-Array. From this table the visibility plot from the Teide observatory has been obtained for each target.  

\begin{table}
\caption{List of selected \gray{} sources relevant for the study of CR origin (Pillar-1) and of cosmology and fundamental physics (Pillar-2), observable from the Observatorio del Teide, adapted from Paper II \citep{2022JHEAp...S..XXXS}.}
\centering
\def\arraystretch{1}
\begin{tabular}{L@{\hskip 0.05in}R@{\hskip 0.05in}R@{\hskip 0.05in}C@{\hskip 0.05in}C@{\hskip 0.05in}}
\toprule
Name & RA & Dec & Type  & Category \\
 & (deg) & (deg) &  & \\
\midrule
Tycho                &   6.36 & +64.13 & SNR          & Pillar 1 \\
Galactic Center      & 266.40 & -28.94 & Diffuse      & Pillar 1 \\ 
VER J1907$+$062      & 286.91 &  +6.32 & SNR+PWN      & Pillar 1 \\
SNR G106.3+2.7       & 337.00 & +60.88 & SNR          & Pillar 1 \\
$\gamma$-Cygni       & 305.02 & +40.76 & SNR          & Pillar 1 \\
W28/HESS\,J1800-240B & 270.11 & -24.04 & SNR/MC       & Pillar 1 \\
Crab                 &  83.63 & +22.01 & PWN          & Pillar 1 \\
Geminga              &  98.48 & +17.77 & PWN          & Pillar 1 \\
M82                  & 148.97 & +69.68 & Starburst    & Pillar 1 \\
\midrule
IC 310               &  49.18 & +41.32 & Radio galaxy & Pillar 2 \\
M87                  & 187.70 & +12.40 & Radio galaxy & Pillar 2 \\
Mkn 501              & 253.47 & +39.76 & Blazar       & Pillar 2 \\
\bottomrule
\end{tabular}
\label{tab:Pillars}
\end{table}

\begin{figure}
	\centering
		\includegraphics[scale=0.6]{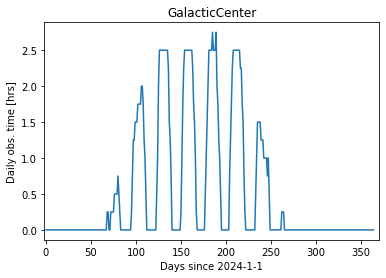}
	\caption{Visibility of the Galactic Center from the Teide observatory for the year 2024.}
	\label{FIG:gal_center}
\end{figure}

\begin{figure}
	\centering
		\includegraphics[scale=0.6]{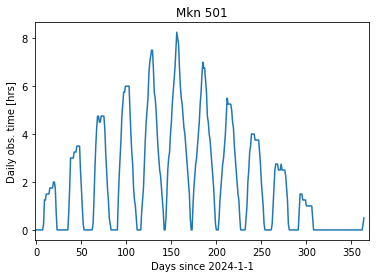}
	\caption{Visibility of Mkn 501 from the Teide observatory for the year 2024.}
	\label{FIG:Mkn501}
\end{figure}

\begin{figure}
	\centering
		\includegraphics[scale=0.55]{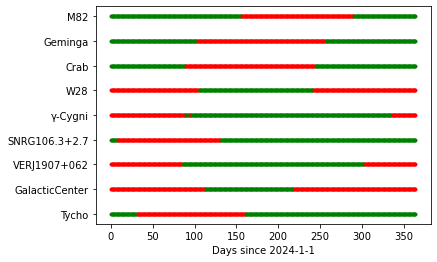}
	\caption{Visibility periods in 2024 for all the targets in table \ref{tab:Pillars}. Green indicates nights with more than 2 hours of available observing time.}
	\label{FIG:visibility}
\end{figure}

\begin{figure}
	\centering
		\includegraphics[scale=0.54]{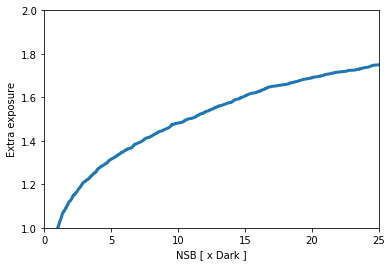}
	\caption{Increase of the AAOT as a function of the maximum NSB (Night Sky Background) allowed for observations. The slope of this curve depends on the source position, here it is shown as obtained for the Crab nebula.}
	\label{FIG:Crabmoonlight}
\end{figure}
The results of this analysis is shown in Figures \ref{FIG:gal_center}, and \ref{FIG:Mkn501} which give the visibility throughout the year 2024 for the galactic center and Mkn 501, respectively. Figure \ref{FIG:visibility}, instead, gives the overall picture of the visibility plots for all the targets in Table \ref{tab:Pillars} that can give useful information to implement an observing strategy. For example, starting from these visibility plots, one can infer that the Galactic center is visible approximately 138 hr/yr or that Mkn 501 is visible 144.5 hr/yr. To obtain these numbers no attempt was done to optimize the observation strategy through a carefully scheduling of the observing nights. We estimated that, in three years of operations, we should be able to devote at least 1500 observing hours to the first “Pillar science topic” (i.e. origin of Cosmic Rays, see paper II) and about 1000 hours to the second “Pillar science topic” (Cosmology and fundamental physics). These numbers refer to only the dark (moonless) time and they have to be compared to the results of the simulations reported in paper II that for each source give the number of observing hours necessary to obtain significant scientific results. However the ASTRI camera can take data also with medium-high levels of Night Sky Background (up to about 15 times the dark NSB).  This will allow the observation with moderate moonlight conditions, increasing significantly the AAOT  (Figure \ref{FIG:Crabmoonlight}). In these conditions we expect to have little degradation of the sensitivity, especially at high energies, as reported for other Cherenkov telescopes (see for example: Archambault et al., \citeyear{veritasMoonlight}, Ahnen et al.,  \citeyear{magicMoonlight}.

\section{Conclusions}
The ASTRI Mini-Array is an INAF project to build and operate nine innovative Imaging Atmospheric Cherenkov telescopes. Thanks to an agreement between INAF and IAC the telescopes will be installed at the Teide Astronomical Observatory. The facility will operate for at least 8 years 
The ASTRI mini-array will be the largest facility of IACT arrays until the CTA observatory will start operations. Thanks to its expected overall performance, the ASTRI MA will represent an important instrument to perform deep observations of the Galactic and extra-Galactic sky at these energies. The current master schedule foresees the full array ready for commissioning by mid of 2024 with the operations starting six months later. 

\section*{Acknowledgments}
\noindent
This work was conducted in the context of the ASTRI Project. This work is supported by the Italian Ministry of Education, University, and Research (MIUR) with funds specifically assigned to the Italian National Institute of Astrophysics (INAF). We acknowledge support from the Brazilian Funding Agency FAPESP (Grant 2013/10559-5) and the South African Department of Science and Technology through Funding Agreement 0227/2014 for the South African Gamma-Ray Astronomy Programme. This work has been supported by H2020-ASTERICS, a project funded by the European Commission Framework Programme Horizon 2020 Research and Innovation action under grant agreement n. 653477. IAC is supported by the Spanish Ministry of Science and Innovation (MICIU).

The ASTRI project acknowledges the fundamental contribution of several industrial partners that are developing and producing some of the main subsystems. In particular we  are grateful to EIE group, Media Lario, ZAOT, CAEN, Nuclear Instruments, Weeroc, Hamamatsu Photonics Italia for their very valuable support. 

The ASTRI project is becoming a reality thanks to Giovanni ``Nanni'' Bignami, Nicol\`{o} ``Nichi'' D'Amico two outstanding scientists who, in their capability of INAF Presidents,  provided continuous support and invaluable guidance. While Nanni was instrumental to start the ASTRI telescope, Nichi transformed it into the Mini Array in Tenerife. Now the project is being built owing to the unfaltering support of Marco Tavani, the new current INAF President. Paolo Vettolani and Filippo Zerbi, the past and current INAF Science Directors, as well as Massimo Cappi, the Coordinator of the High Energy branch of INAF, have been also very supportive to our work. We are very grateful to all of them. Nanni and Nichi, unfortunately, passed away but their vision is still guiding us.




\bibliographystyle{apsrev}

\bibliography{MiniArray}

\vskip 1pt

\bio{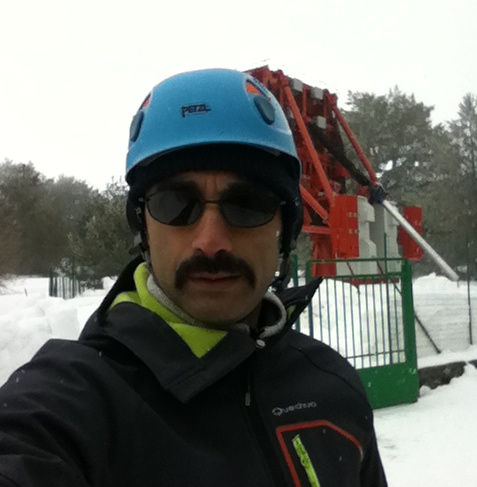}
Salvatore Scuderi got his master degree in physics in 1989 at the University of Catania and then obtained the PhD in 1994 working on the properties of stellar winds of early type stars. ESA fellow at the Space Telescope Science Institute from March 1994 until March 1996. From 1995 to 2018 has been staff astronomer at the INAF OA Catania and then moved to INAF IASF Milan. Its main activity is the development, realization, and testing of instrumentation for ground based and space observatories. He has been involved in several national and international projects (TNG-SARG, TNG-GIANO,UVISS, WSO-UV, VLT-SPHERE, EST, CHEOPS, PLATO, SOXS, ASTRI, CTA) where has played different roles, WP leader, System Engineer, Project Manager. He is currently the Project Manager of ASTRI Mini-Array project.
\endbio

\end{document}